\documentclass[a4paper,fleqn,usenatbib]{mnras}

\usepackage[T1]{fontenc}
\usepackage{ae,aecompl}

\usepackage{graphicx}	
\usepackage{amsmath}	
\usepackage{amssymb}	

\title[Two-Temperature Accretion Flow Around Black Holes]{Properties of Two-Temperature
Dissipative Accretion Flow Around Black Holes}

\author[I. K. Dihingia et al.]{
Indu K. Dihingia,$^{1}$\thanks{E-mail: i.dihingia@iitg.ernet.in}
Santabrata Das$^{1}$\thanks{E-mail: sbdas@iitg.ernet.in}
and Samir Mandal$^2$\thanks{E-mail: samir@iist.ac.in}
\\
$^{1}$Indian Institute of technology Guwahati, Guwahati, 781039, India\\
$^2$Indian Institute of Space Science and Technology, Thiruvananthapuram, India
}

\date{Accepted XXX. Received YYY; in original form ZZZ}

\pubyear{2016}

\begin{document}
\label{firstpage}
\pagerange{\pageref{firstpage}--\pageref{lastpage}}
\maketitle

\begin{abstract}

We study the properties of two-temperature accretion flow around a non-rotating
black hole in presence of various dissipative processes where pseudo-Newtonian potential
is adopted to mimic the effect of general relativity.
The flow encounters energy loss by means of radiative processes acted on the electrons
and at the same time, flow heats up as a consequence of viscous heating effective on ions. We
assumed that the flow is exposed with the stochastic magnetic fields which leads to Synchrotron
emission of electrons and these emissions are further strengthen by Compton scattering. We
obtain the two-temperature global accretion solutions in terms of dissipation parameters, namely, 
viscosity ($\alpha$) and accretion rate (${\dot m}$), and find for the first time in the literature
that such solutions may contain standing shock waves. Solutions of this kind are multi-transonic
in nature as they simultaneously pass through both inner critical point ($x_{\rm in}$) and
outer critical point ($x_{\rm out}$) before crossing the black hole horizon. We calculate
the properties of shock induced global accretion solutions in terms of the flow parameters.
We further show that two-temperature shocked accretion flow is not a discrete solution,
instead such solution exists for wide range of flow parameters. We identify the effective 
domain of the parameter space for standing shock and observe that parameter space shrinks
as the dissipation is increased. 
Since the post-shock region is hotter due to the effect of shock compression, it naturally 
emits hard X-rays and therefore, the two-temperature shocked accretion solution has the
potential to explain the spectral properties of the black hole sources.

\end{abstract}


\section{Introduction}

The accretion of matter around black hole is considered to be the key physical
mechanism both in X-ray binaries (XRBs) and active galactic nuclei (AGNs) as 
immensely  large amount of energy is released in this processes 
\citep{Frank2002}. In order to explain the observed hard X-ray spectra from 
black hole candidate Cyg-X1, \citet[]{SLE1976} (hereafter SLE) suggested 
two-temperature  accretion disc model  as standard \citet*{Shakura_sunyav1973}
(hereafter SS73) disc was unable to explain observed hard X-rays. Very soon, 
the above approach drew lot of attention among the  accretion physicists due to the 
presence of a `hotter' branch in the solution of SLE model. Eventually, 
SLE model has been widely studied and applied to XRBs and AGNs (e.g. 
\cite{kusunose1988two,kusunose1989two,white1989hot,
wandel1991hybrid,luo1994stability}). However, \citet*{pringle1976thermal} and
\citet*{piran1978role} showed that SLE model is thermally unstable. The growth 
of instability in SLE model is so rapid that may brake the equilibrium of the 
accretion disc, which is not expected from the realistic configuration of an 
accretion disc \citep{shi2000radial}. So, advection comes into the picture of 
SLE model. The effect of advection on SLE model has been studied extensively 
for black hole accretion 
\citep{abramowicz1988,kato1988viscous,Narayan_pophem1993,abramowicz1994thermal,
Narayan1994,Narayan-Yi95,chen1995hot,chen1994variability,misra1996crucial,
nakamura1996global,wu1997stability,shi2000radial}. It has been established 
that thermal instability exists only if the disc is cooling dominated but 
disappears with addition of advection in it. Similarly, SS73 model is also 
appeared unstable due to the thermal and viscous instabilities
\citep{Lightman1974,Eardley1975,shakura1976}.
Therefore, based on the above considerations, a two-temperature viscous 
accretion disc model is invoked
as a realistic configuration of accretion disc around black holes
\citep{Narayan-Yi95,nakamura1996global,Chakrabarti_Titarchuk95,
manmoto1997spectrum,Mandal2005,Rajesh2010}.

Initially, \citet{Narayan-Yi95} investigated two-temperature accretion flow around
black hole including radial advection, particularly for low mass accretion rate.
In their study, they adopted the method of self-similarity while obtaining the
global accretion solutions. Latter on, several attamepts were made 
to calculate the self-consistent generalized transonic accretion solutions of
two-temperature collisionles plasmas accreting on to compact objects
\citep{nakamura1996global,manmoto1997spectrum,Rajesh2010}.
However, in all these studies, special emphasis was given on a specific class
of accretion solutions that only pass through the inner critical points before crossing
the horizon. Indeed, these solutions are merely a subset of complete global accretion solutions
around black holes.
Meanwhile, \citet{Chakrabarti_Titarchuk95,Mandal2005,ChakrabartiMandal2006} 
investigated the feasibility of
two-temperature accretion solutions including shock waves while modeling the spectral
characteristics commonly observed in black hole sources. Nevertheless, these models 
were developed considering the fact that the self-consistent electron heating
effect is decoupled from the governing hydrodynamical disc equations.   

In an accretion disc around black hole, rotating
flow experiences centrifugal repulsion against gravity that effectively
constructs a virtual barrier in the vicinity of the black holes. Depending
on the flow variables, eventually, such a virtual barrier triggers discontinuous
transition of the flow variables in the form of shock waves. 
Indeed, \citet{Becker-Kazanas01} argued that according to the second law of
thermodynamics, shock induced global accretion solutions are thermodynamically
preferred as they possess high entropy content.
Meanwhile, several attempts were made in order to examine the existence of shock
waves in the accretion flows around black holes both in theoretical as well as 
numerical fronts \citep{Fukue87,Chakrabarti1989,Molteni1995,Chakrabarti96,Molteni1996a,
Lu1999,Becker-Kazanas01,Das2001a,Chakrabarti2004,Fukumura2004,Das2007,Das2014,Okuda2015,
aktar2015,Sarkar-Das16,Aktar-etal17,Sarkar-etal17}. Due to shock compression, since the post-shock flow
becomes hot and dense, all the relevant thermodynamic and radiative
processes are very much active in this region that eventually reprocessed the
synchrotron and bremsstrahlung photons in the post-shock region
into hard radiations via inverse
Comptonization mechanism \citep{Chakrabarti_Titarchuk95,Mandal2005}. Apparently,
the presence of shock waves in the accretion solutions leads to the
formation of post-shock corona (PSC) around the black holes 
and therefore, PSC
seems to regulate the spectral features of the black hole sources. Moreover,
due to presence of excess thermal gradient across the shock front, a part
of the accreting matter is deflected at PSC and eventually diverted in the
vertical direction to produce bipolar jets and outflows \citep{Chakrabarti99,
Das2001b,Das_Chattopadhyay08,aktar2015}. Interestingly, when PSC undulates, the
emergent hard radiations exhibit quasi-periodic oscillation (QPO) which is
observed in spectral states of many black hole sources 
\citep{Chakrabarti_Manickam00,Nandi_etal01a,Nandi_etal01b,Nandi_etal12}.
Meanwhile, with an extensive numerical study, \citet{Das2014} showed that
the oscillation of PSC displays QPO features along with the variable mass
outflows originated from the inner part of the disc. 

It is to be noted that all the above studies were carried out considering
the fact that the electrons and ions hold very strong coupling between
them and hence, they are able to maintain a common single temperature
profile all throughout.
Very recently, in a brief note, \citet*{dihingia2015shocks} reported that
two-temperature global accretion solutions may harbour shock waves and
subsequently hinted that solutions of this kind are very much promising
to explain the characteristics of the spectral states of the black hole
sources as well.  
Motivating with this, in this paper, we model the two-temperature accretion
flow around a Schwarzschild black hole in the realm of sub-Keplerian flow paradigm.
We consider a set of governing equations that describe the accreting matter
in presence of various dissipative processes, such as viscosity and radiative cooling.
We adopt a pseudo-Newtonian
potential to mimic the space-time geometry around the black hole 
\citep{Paczynsky_wiita1980}. This eventually allow us to tackle the problem  
following the Newtonian approach while retaining the salient
features of general relativity. We consider bremsstrahlung, synchrotron
and Compton cooling mechanisms as key radiative processes active in the flow. Moreover,
in order to address the heating of ions due to viscous dissipation and the angular
momentum transport of the accretion flow, we regard
the mixed shear stress prescription to delineate the effect of viscosity as proposed by
\citet{Chakrabarti_Molteni1995}. With this, we self-consistently solve the
governing equations to obtain two-temperature global accretion solutions and 
show how the nature of the accretion solution changes with the dissipation
parameters, namely viscosity ($\alpha$) and accretion rate (${\dot m}$).
We identify the accretion solutions that are essential for standing shock
formation and calculate all the relevant dynamical and thermodynamical flow
variables for shocked accretion solutions. We observe that dynamics of the
shock front (equivalently PSC) can be controlled by the dissipation parameters,
in particular, shock moves toward the horizon when strength of dissipation
is increased. Further, we examine the shock properties in terms of the
dissipation parameters and find that shocked accretion solution is not an
isolated solution, instead such solution exists even for
high dissipation limits. Moreover, we identify the effective region of
parameter space spanned by the specific energy (${\mathcal E}_{\rm in}$)
and specific angular momentum ($\lambda_{\rm in}$) of the flow measured
at the inner critical point ($x_{\rm in}$) and find that parameter space
shrinks when the dissipation is increased. Overall, in this work, we
demonstrate a useful formalism to study the two-temperature global 
accretion flow including shock wave that provides self-consistent accretion
solutions for a wide range of viscosity ($\alpha$) and accretion rate (${\dot m}$).
To our knowledge, no self-consistent theoretical attempt 
has been made so far 
to study the two-temperature global transonic accretion flow including
shock wave by combing the hydrodynamics and radiative processes 
simultaneously.

In \S 2, we describe the governing equations and perform 
the critical point analysis.
In \S 3, we present the global accretion solutions with and without
shock waves and study the shock dynamics and shock properties as a function
of dissipation parameters. In \S 4, we classify the shock parameter space
and finally, in \S 5, we present discussion and concluding remarks.

\section{Assumptions and Model Equations}

\label{sec:maths}
\subsection{Basic Hydrodynamics}

In order to investigate the hydrodynamic properties of the two-temperature 
accreting plasma, we consider a steady, thin, 
axisymmetric and viscous accretion disc around a Schwarzschild black hole. 
While doing this, we choose a cylindrical coordinate system where the
central black hole is assumed to be located at the origin. The space-time
geometry around the black hole is approximated by adopting the pseudo Newtonian
potential $\Phi = -GM_{\rm BH}/(r-2GM_{\rm BH}/c^2)$ \citep{Paczynsky_wiita1980},
where, $G$ is the gravitational constant, $M_{\rm BH}$ is the mass of the
black hole, $c$ is the speed of light and $r$ is the radial coordinate. 
In this work, we use an unit system as $G=1/2$, $M_{\rm BH} =1$ and $c=1$, respectively
that allows us to express all the flow variables in dimensionless unit. 
With this, the units of radial coordinate, velocity and specific
angular momentum of the flow are expressed as  
$2GM_{\rm BH}/c^2$, $c$ and $2GM_{\rm BH}/c$, respectively. For representation,
in this work, we choose the mass of the black hole as $10M_{\odot}$ all throughout
unless stated otherwise, where $M_{\odot}$ stands for the solar mass.

In the steady state, the hydrodynamic equations that govern the infalling matter
are given by,

\noindent (a) The radial momentum equation:
$$\label{eq:2}
u\frac{du}{dx}+\frac{1}{\rho}\frac{dP}{dx}-\frac{\lambda^2}{x^3} + F(x) = 0,
\eqno(1)
$$
where, $x$ is the radial distance, $u$ is the radial velocity, $\lambda$ is the
specific angular momentum, $P$ is the isotropic pressure and $\rho$ is the
density of the flow, respectively. Moreover, $F(x)$ denotes 
the gravitational acceleration and is obtained as $F(x) =d\Phi/dx$.

\noindent (b) The mass conservation equation:

$$\label{eq:1}
\dot{M}=2\pi\Sigma x u,
\eqno(2)
$$
where, $\dot{M}$ represents the mass accretion rate which we treat as global 
constant all through out. Moreover, in this work, we consider
$u$ to be positive. In the subsequent analysis, we express accretion rate 
in terms of Eddington rates as ${\dot m} = {\dot M}/{\dot M}_{\rm Edd}$. 
Here,  $2\pi$ is the geometric constant and $\Sigma$
denotes the surface mass density of the accreting matter \citep{Matsumoto1984}. 

\noindent (c) The azimuthal momentum equation:
$$\label{eq:3}
u\frac{d\lambda}{dx}+\frac{1}{\Sigma x}\frac{d}{dx}\left(x^2W_{x\phi}\right)=0.
\eqno(3)
$$
Here, we assume that $x-\phi$ component of the viscous stress dominates 
over rest of the components and it is denoted by $W_{x\phi}$.

\noindent (d) Finally, the entropy equations for ions and electrons are given by,
$$\label{eq:4}
\frac{u}{\gamma_i-1}\left(\frac{1}{\rho_i}\frac{dP_i}{dx}-\frac{\gamma_i P_i}
{\rho_i^2}\frac{d\rho_i}{dx}\right)=\Lambda_i-\Gamma_i,
\eqno(4)
$$
and
$$\label{eq:5}
\frac{u}{\gamma_e-1}\left(\frac{1}{\rho_e}\frac{dP_e}{dx}-\frac{\gamma_e P_e}
{\rho_e^2}\frac{d\rho_e}{dx}\right)=\Lambda_e-\Gamma_e,
\eqno(5)
$$
where, $\gamma_{i, e}$ represents the ratio of specific heats. In addition,
$\Gamma_{i, e}$ is the dimensionless total heating terms and $\Lambda_{i, e}$ is the dimensionless 
total cooling terms of the ions and electrons, respectively.
We consider $\gamma_i = 3/2$ for non-relativistic ions as the thermal energy
of ions usually never exceeds $10\%$ of the rest mass energy of ions and
$\gamma_e = 4/3$ for relativistic electrons. 
Moreover, total gas pressure ($P$) of the flow is given by,
$$
P = P_i + P_e = \frac{\rho_i k_B T_i}{\mu_i m_p} + \frac{\rho_e k_B T_e}{\mu_e m_e},
$$
where, the effect due to radiation pressure is neglected. Here,
the quantities have their usual meaning with subscripts $i$ and $e$ are for
ions and electrons, respectively.
In addition, $m_e$ and $m_p$ denote the mass of the electron and ion,
$k_B$ is the Boltzmann constant and $\mu$ is the mean molecular weight
which is given by $\mu_{i}=1.23$ and $\mu_{e}=1.14$ \citep{Narayan-Yi95}.
With this, in this work, we define the sound speed as $a=\sqrt{P/\rho}$.

Following \citet{Chakrabarti_Molteni1995}, we adopt the viscosity prescription 
as $W_{x\phi} =-\alpha (W + \Sigma u^2)$ where, $\alpha$ refers the viscosity 
parameter and $W$ is the vertically integrated gas pressure \citep{Matsumoto1984}. This
provides the heating of ions due to viscous dissipation as,
$$
\Gamma_i = -2\alpha I_n x\left(g a^2 + u^2\right)\frac{d\Omega}{dx}.
\eqno(6)
$$
where, $g = I_{n+1}/I_n$,  $I_n = (2^n n!)^2/(2n+1)!$ 
\citep{Matsumoto1984}, $n = 1/(\gamma_i -1)$ 
and $\Omega$ is the angular velocity of the flow. The cooling of 
ions takes place due to the energy transfer from ions to electron 
via Coulomb coupling $(Q_{ei})$ and through the inverse bremsstrahlung
$(Q_{ib})$ process. The overall cooling rate of ions is therefore given by,

$$
Q_i^- = Q_{ei} + Q_{ib},
\eqno(7)
$$
where, the explicit expression of the Coulomb coupling is 
given by \citep{Spitzer2013,Colpi1984,Mandal2005}:
$$
Q_{ei} = 1.6\times10^{-13}\frac{K_B\sqrt{m_e}{\rm ln}~\Lambda_0}{m_p}n^2
\left(T_i - T_e\right){T_e}^{-3/2},
\eqno(8)
$$
and the inverse bremsstrahlung is given by \citep{Rybicki1979,Colpi1984,Mandal2005}: 
$$
Q_{ib} = 1.4\times10^{-27}n^2\left(\frac{m_e}{m_p}T_i\right)^{1/2}.
\eqno(9)
$$
In Eq. (8), ${\rm ln}~\Lambda_0$ is the Coulomb logarithm and in Eq. (8-9), $n$ is
the number density distribution of the flow (same for ion and electron). 
Finally, we obtain the dimensionless total cooling term for ions as 
$\Lambda_{i}=\left(Q_{i}^{-}/\rho\right) \times (2GM_{\rm BH}/c^5)$.

The same coulomb coupling acts as heating terms for electrons and therefore we have,
$$
Q_e^+ = Q_{ei}.
\eqno(10)
$$

Electrons are lighter than ions and cool more efficiently by bremsstrahlung 
process $(Q_b)$, cyclo-synchrotron process $(Q_{cs})$, and Comptonization
process $(Q_{mc})$, respectively. Thus, the effective electron cooling is
obtained as,
$$
Q_e^- = Q_b + Q_{cs} + Q_{mc}.
\eqno(11)
$$
where, the explicit expressions for these terms are given by \citep{Rybicki1979,Mandal2005},\\

\noindent (a) bremsstrahlung process:
$$
Q_b = 1.4\times 10^{-27}n^2T_e^{1/2}\left(1 + 4.4\times10^{-10}T_e\right).
\eqno(12)
$$
\noindent (b) cyclo-synchrotron process:
$$
Q_{cs} = \frac{2\pi}{3c^2}K_BT_e\frac{\nu_{a}^3}{x},
\eqno(13)
$$
where, $\nu_{a}$ is the critical frequency or cut-off frequency of
synchrotron self-absorption, which depends on the
electron temperature and magnetic field. We consider only thermal electrons
and calculate $\nu_a$ by equating the surface and volume emission of 
synchrotron radiation from PSC \citep{Mandal2005}. The expression of $\nu_a$ is 
given by,
$$
\nu_a = \frac{3}{2}\nu_0\theta_e^2x_m,
\eqno(14)
$$
where, 
$$
\theta_e = \frac{K_B T_e}{m_e c^2} ;~~~~
\nu_0 = \frac{eB}{2\pi m_ec} ~~{\rm and} ~~~~x_m = \frac{2\nu_c}{3\nu_0\theta_e^2}.
\eqno(15)
$$

Synchrotron emission is observed when the charged particles
interact with the magnetic fields. 
In the absence of satisfactory description of magnetic fields in an
accretion disc, we assume the existence of random or stochastic magnetic
fields that possibly originate due to the turbulent nature of the
accretion flow. Such magnetic fileds may or may not be in equipartition
with the flow. With this consideration, we approximate the strength of the magnetic
fields $B$ taking the ratio of the magnetic pressure to the gas pressure as
a global parameter ($\beta$) as
$$
\beta = \frac{B^2/8\pi}{P}.
\eqno(16)
$$
In general, $\beta \lesssim 1$ in order to ensure that the magnetic fields
remain confined within the disc. In this work, we consider $\beta=0.1$ all
throughout unless stated otherwise.

\noindent (c) Comptonization process:
$$
Q_{mc} = Q_{cs}\mathcal{F}.
\eqno(17)
$$
Here, $\mathcal{F}$ is the enhancement factor for Comptonization of synchrotron 
radiation. We follow the prescription proposed by \citet*{Mandal2005} to calculate $Q_{mc}$.

Finally, we obtain the 
dimensionless total heating and cooling terms for electrons  as 
$\Gamma_{e}=\left(Q_{e}^{+}/\rho\right) \times (2GM_{\rm BH}/c^5)$ and 
$\Lambda_{e}=\left(Q_{e}^{-}/\rho\right) \times(2GM_{\rm BH}/c^5)$,
respectively.

\subsection{Critical Point Analysis}

To obtain the global accretion solution around a black hole, we solve
Eqs. (1-5) by following the method described in 
\citet{Chakrabarti1989, Das2007}. Evidently, these accretion flows must be transonic
in nature in order to satisfy the inner boundary conditions imposed by the black hole
event horizon. Following the above insight, we proceed to obtain the 
critical point conditions by calculating the radial velocity gradient as,

$$
\frac{du}{dx} = \frac{N}{D},
\eqno(18)
$$
where, the numerator $(N)$ and the denominator $(D)$ are given by,

$$
\begin{aligned}
N = & ~\Lambda_i(x, u, a, \lambda, T_e)\\  
&+\frac{2\alpha^2I_n(ga^2 + u^2)^2}{ux} - 
	\frac{4\alpha I_n(ga^2 + u^2)\lambda}{x^2}\\
 &+ \frac{4\alpha^2I_nga^2(ga^2 + u^2)}{u}
\left\{\frac{3}{2x} + \frac{1}{2F(x)}\frac{dF(x)}{dx}\right\} \\
&- \bigg[\frac{\gamma_i + 1}{\gamma_i - 1}u- \frac{4\alpha^2I_ng(ga^2 + u^2)}{u}
\bigg]\times\bigg\{\frac{\lambda^2}{x^3} - F(x)\bigg\}\\
&-\frac{2\gamma_iua^2}{\gamma_i-1}\left\{\frac{3}{2x} - \frac{1}{2F(x)}\frac{dF(x)}{dx}
\right\},\\
\end{aligned}
\eqno(19)
$$
and

$$
D =\frac{2\gamma_ia^2}{\gamma_i - 1} - \frac{\gamma_i + 1}{\gamma_i - 1}
u^2 + 2\alpha^2I_n(ga^2 + u^2)\bigg\{(2g - 1) - \frac{ga^2}{u^2}\bigg\}.\\
\eqno(20)
$$

It is to be noted that within the range of few tens of Schwarzschild radius ($r_g$) from the
black hole horizon, electron temperature ($T_e$) usually remains lower than the
ion temperature ($T_e$) and therefore, for simplicity, 
we consider $P \sim P_i$ to obtain equation (18).

The gradient of sound speed is calculated as, 
$$\frac{da}{dx} =~\bigg\{\frac{a}{u} - \frac{u}{a}\bigg\}\frac{du}{dx} + 
\frac{1}{a}\bigg\{\frac{\lambda^2}{x^3} - F(x)\bigg\}
 + \bigg\{\frac{3a}{2x} + \frac{a}{2F(x)}\frac{dF(x)}{dx}\bigg\},\\
\label{eq:dadx}
\eqno(21)
$$

The gradient of angular momentum is obtained as,
$$
\frac{d\lambda}{dx} =~\frac{\alpha(ga^2 + u^2)}{u} - \frac{\alpha x (ga^2 - u^2)}
{u^2}\frac{du}{dx} + \frac{2\alpha gax}{u}\frac{da}{dx},
\label{eq:dldx}
\eqno(22)
$$

The gradient of electron temperature is given by,
$$
\begin{aligned}
\frac{dT_e}{dx} =&~\frac{(\gamma_e -1)m_p\mu_e}{k_{B}u}(\Lambda_e - \Gamma_e)\\
 - &(\gamma_e - 1)T_e\bigg\{\frac{3}{2x} + \frac{1}{2F(x)}\frac{dF(x)}{dx}\bigg\}
-\frac{(\gamma_e - 1)T_e}{a}\frac{da}{dx}\\
& - \frac{(\gamma_e - 1)
T_e}{u}\frac{du}{dx}.\\
\end{aligned}
\label{eq:dtdx}
\eqno(23)
$$

In reality, the flow starts accreting with negligible radial velocity from
the outer edge of the disc and eventually, crosses the event horizon with
velocity equal to the speed of light. This findings demand that the radial
velocity gradient of the flow along the streamline must be real and finite
everywhere between the outer edge of the disc and the event horizon.
However, Eq. (20) suggests that there may be locations where the denominator
($D$) vanishes. In order to maintain the flow to be smooth everywhere
along the streamline, the numerator must also vanishes at the same point
where $D$ goes to zero. Such special points where both $N$ and $D$
simultaneously vanish are known as the critical points and
the conditions $N=D=0$ are called as critical point conditions. Setting
$D=0$, we find Mach number ($M=u/a$) at the critical point ($x_c$)
which yields as, 
$$
M_{c}= \left(\frac{-m_{2} \pm \sqrt{m_{2}^{2} - 4 m_{1} m_{3}}}{2m_{1}}\right)^{1/2}
\eqno(24)
$$
where,
$$
\begin{aligned}
m_1 = &(\gamma_i + 1) - 2\alpha^2g(2g-1)(\gamma_i-1),\\
m_2 = &4\alpha^2g(1-g)(\gamma_i -1) - 2\gamma_i,\\
m_3 = &2\alpha^2g^2(\gamma_i -1).\\
\end{aligned}
$$

Now, setting $N=0$, we obtain the algebric equation of the sound speed which is 
given by,

$$
\begin{aligned}
\Lambda_i(x_c,a_c,\lambda_c,T_{e, x_c}) + 
\mathcal{A}a^3_c +
\mathcal{B}a^2_c + \mathcal{C}a_c = 0,
\end{aligned}
\eqno(25)
$$
where,
$$
\mathcal{A} = \frac{2\alpha^2I_n(g + M_c^2)^2}{M_c x_c}
$$
$$
+ \left\{\frac{4\alpha^2I_ng(g+M_c^2)}{x_c M_c} - \frac{2\gamma_i M_c}
{\gamma_i-1}\right\} \left\{\frac{3}{2x_c} + \frac{1}{2F(x_c)}\frac{dF(x)}{dx}\bigg\rvert_{c} \right\},
$$
$$
\mathcal{B} = -\frac{4\alpha I_n(g+M_c^2)\lambda_c}{x_c^2},
$$
$$
\mathcal{C} = 
	-\left\{\frac{\lambda_c^2}{x_c^3} - F(x_c)\right\}
	\left\{\frac{\gamma_i + 1}{\gamma_i - 1}M_c - \frac{4\alpha^2I_ng(g + M_c^2)}
{M_c}\right\}.
$$

We solve equation (25) to find the sound speed at the critical point ($a_c$) using the
input parameters of the flow. Subsequently, we calculate the radial velocity at the 
critical point ($u_c$) using equation (24).
At the critical point, radial velocity gradient has the form $(du/dx)=0/0$,
and therefore, we apply the l$^{\prime}$Hospital rule to calculate the radial velocity
gradient at the critical point ($x_c$). Details are given at Appendix. It is to be noted 
that the linear expansion of the wind equation (18) around the critical points
also provides the same result \citep{Kopp_Holzer_1976,Holzer_1977,Jacques_1978}.
In general, at the critical point, radial velocity gradient 
owns two distinct values;
between them one is for accretion and the remaining one is for wind. When both
the radial velocity gradients are real and of opposite sign, the corresponding
critical point is called as saddle type critical point \citep[and references therein]{Matsumoto1984,
Kato-etal1993,Chakrabarti2004} and such critical points have special importance as the global
transonic accretion flow can only pass through it. Accordingly,
in this work, we deal with the saddle type critical points
which are subsequently referred as critical points
in rest of the paper. Interestingly,
accretion flow may contain multiple critical points depending on the input 
parameters. When critical points form close to the horizon, we call them 
as inner critical points ($x_{\rm in}$) and when they form far away from the
black hole, we call them as outer critical points ($x_{\rm out}$).
The procedure to obtain the location of multiple critical points is explicitly
described in the next section.

\section{Global Accretion Solutions}

Following the above considerations, we obtain the two-temperature global 
accretion solution around black hole by employing the flow parameters assigned 
at a given radial coordinate. Evidently, critical point location turns out to be
the preferred radial coordinate for the same as  
all the physically acceptable black hole solutions
are transonic in nature that connects the black hole horizon with the
outer edge of the disc. Accordingly, we supply the following input parameters,
namely, $T_{e,c}$, $\lambda_c$, $\alpha$, and ${\dot m}$,
respectively at the critical point location ($x_c$) as the boundary conditions and
integrate equations (18-23) from $x_c$, once inward up to the horizon and 
then outward up to the outer edge of the disc. Finally, we join these two 
parts of the solution in order to get a complete transonic global accretion 
solution. 

\subsection{Accretion solutions containing inner critical point}

\begin{figure} 

\includegraphics[width=0.485\textwidth]{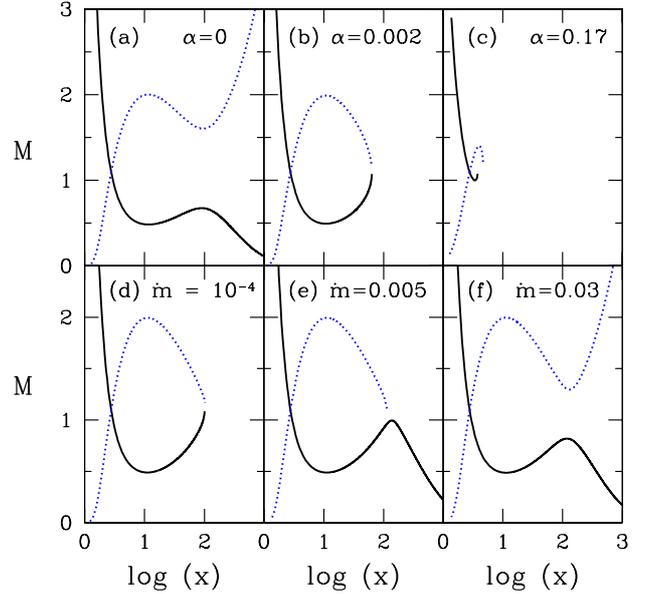}
\caption{Variation of Mach number $M(=u/a)$ with radial distance $x$. 
Solid curve denotes the accretion solution and dotted curve
represent the corresponding wind solution. See text for details.}
\label{fig:fig_01}
\end{figure}

In Fig. 1, we investigate the variation of Mach number ($M$) of the flow as
function of logarithmic radial distance. In the upper panel, we examine the
effect of viscosity ($\alpha$) on the transonic accretion solution, while
in the lower panel, effect of accretion rate (${\dot m}$) is studied. In
Fig. 1a, we choose the input parameters of the flow at the inner critical point
as $x_{\rm in}=2.784$, $\lambda(x_{\rm in})=1.543$ (hereafter, $\lambda_{\rm in}$),
$T_e(x_{\rm in})= 8.7\times10^9{\rm K}$ (hereafter, $T_{\rm e,~in}$), $\alpha=0$, and
${\dot m} = 0.005$, respectively. In this work, although we intend to focus only on the 
accretion solutions, in Fig. 1, we illustrate both accretion (solid curve) and its corresponding
wind branch (doted) for the purpose of completeness.
In the invicid limit,
the sub-sonic flow at the outer edge gains its radial velocity and subsequently
becomes supersonic after crossing the inner critical point before falling in to
the black hole. 
Solutions of this kind was studied earlier by 
\cite{Narayan-Yi95,manmoto1997spectrum,Rajesh2010}. In panel (b) and (c), viscosity
is increased further and the accretion solution becomes closed and it fails to
connect the horizon with the outer edge of the disc. This essentially indicates that
there exists a limiting viscosity that changes the character of the
accretion solutions from open to close one. The accretion solution of this kind 
may join with another solution that passes through the outer critical point
provided a shock is formed (this will be discussed in the subsequent section)
and finally merges with the Keplerian disc
at the outer edge.
When the viscosity is increased further, the closed topology having
same inner critical point properties
shrinks gradually and finally disappears. Therefore, when a global accretion
flow possesses multiple critical points, there exists two critical $\alpha$ 
corresponding to a given set of flow parameters, where multiple critical points
induct with the lower critical value of $\alpha$ and then multiple critical points
disappear for the higher critical $\alpha$.

In the lower panel of Fig. 1, we demonstrate the variation of accretion
solutions with accretion rate (${\dot m}$). 
For a very low accretion rate,
we get a closed accretion solution shown in Fig. 1d. In this case,
we choose the input flow parameters same as in Fig. 1a except the viscosity 
as $\alpha = 0.001$. We have already pointed out above that this kind of
solutions do not represent accretion solution unless they are connected
with another solutions passing through the outer critical point via shock
waves. When the accretion rate is increased as ${\dot m}=0.005$ keeping the 
input parameters of the flow unchanged, shock free Bondi-type accretion
solution is obtained which is shown in Fig. 1e. With the continuous
increase of the accretion rate, accretion solution monotonically 
opens up further as depicted in Fig. 1f. Note that there exists a
critical accretion rate that separates the closed accretion solutions
from the open accretion solutions (similar to Bondi-type). Obviously, this critical
${\dot m}$ is not universal, instead depends on the input parameters of
the flow.

\subsection{Accretion solutions with fixed outer edge}

\begin{figure} 
\includegraphics[width=0.485\textwidth]{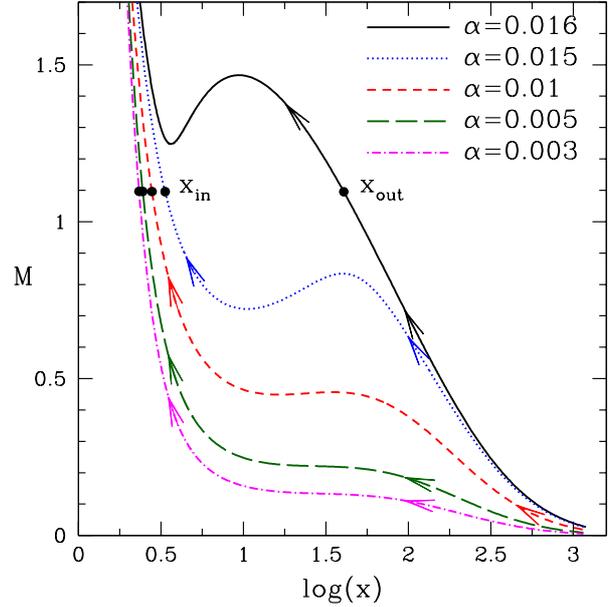}
\caption{Variation of Mach number $M(=u/a)$ with radial distance $x$
for accretion flows injected from a fixed outer edge having different 
viscosity ($\alpha$). Results depicted using dot-dashed, long-dashed,
dashed, dotted and solid curves are for $\alpha = 0.003, 0.005, 0.01, 0.015$
and $0.016$ respectively. Inner and outer critical points are marked in the 
figure as $x_{\rm in}$ and  $x_{\rm out}$. See text for details.}
\label{fig:fig_02}
\end{figure}

During accretion, sub-Keplerian matter deviates from the Keplerian disc
at a large distance and starts accreting towards the black hole subsonically.
In accordance to this accretion scenario, it is worthy to
study the dynamical properties of the accreting matter by fixing the initial 
conditions of the flow at the outer edge of the disc ($x_{\rm edge}$).
Accordingly, for representation, we choose $x_{\rm edge} =1200$ and 
investigate the characteristics of the accretion solution in terms of
the flow parameters fixed at $x_{\rm edge}$. We proceed to select the flow
parameters at $x_{\rm edge}$ in the following manner. 
As before, we consider the input parameters of the flow at the inner 
critical point as $x_{\rm in}=2.329$, $\lambda_{\rm in}=1.719$, 
$T_{\rm e,~in}=1.789\times10^{10}{\rm K}$, $\alpha=0.003$, and ${\dot m}=0.001$,
respectively. Using these parameter  values, we integrate equations (18-23) first inward up 
to horizon and then outward upto $x_{\rm edge} = 1200$.
Subsequently, we join these two parts of the solution to get a complete
transonic global accretion solution which is denoted by the dot-dashed curve 
(magenta) in Fig. 2. Here, we note the flow variables at 
$x_{\rm edge}$ which are obtain as $u_{\rm edge}=1.56\times10^{-4}$,
$a_{\rm edge}=2.798\times10^{-2}$, $\lambda_{\rm edge}=18.827$
and $T_{\rm e,~edge}=4.074\times10^{9}{\rm K}$, respectively.
Moreover, employing these local flow variables, we calculate the local 
energy of the flow at $x_{\rm edge}$ as ${\cal E}_{\rm edge}=2.057\times10^{-3}$
by utilizing the expression of local energy as given by
${\cal E} (x) = 0.5u^2 + \gamma_i a^2/(\gamma_i - 1) + 0.5\lambda^2/x^2 - 0.5/(x-1)$.
In reality, upon integrating equations (18-23) using these outer edge flow 
variables, one can get the same solution.  
Here, arrows indicate the direction of the flow towards the black hole.
Next, we increase viscosity as $\alpha=0.005$ keeping $\lambda_{\rm edge}$,
${\cal E}_{\rm edge}$, $T_{\rm e,~edge}$ and ${\dot m}$ fixed 
at $x_{\rm edge} = 1200$ and obtain another global 
accretion solution by suitably adjusting the values of 
$u_{\rm edge}=2.58\times10^{-4}$ and $a_{\rm edge}=2.800\times10^{-2}$ that 
enters into the black hole after passing through the inner
critical point at $x_{\rm in}=2.4515$. In this case, the additional information of
$u_{\rm edge}$ and $a_{\rm edge}$ is required to obtain the accretion solution as
the critical point is not known apriori. 
In Fig. 2, we plot this solution using long-dashed (green) curve.
Upon increasing the viscosity ($\alpha$) gradually, we identify the maximum value
of viscosity as $\alpha^{\rm max}=0.015$, above this value accretion flow fails to pass
through the inner critical point which is indicated by the dotted (blue) curve.
The obtained results as respectively depicted by dot-dashed, long-dashed, dashed and
dotted curves correspond to the solutions of advection 
dominated accretion flow \citep{Narayan-Yi95,manmoto1997spectrum,Rajesh2010}.
When $\alpha$ is increased further as $0.016$, we find that the accretion solution passes through
the outer critical point ($x_{\rm out}=40.594$) instead of the 
inner critical point ($x_{\rm in}$) which is shown by the solid (black) curve. To our knowledge,
in the context of two-temperature accretion flow, self-consistent accretion 
solutions of this kind are not studied so far. 
Interestingly, accretion flows that become transonic after passing through the
outer critical point ($x_{\rm out}$) are specially portentous as they may
contain centrifugally supported shock waves (see \S 3.3). 

\subsection{Procedure of calculating shock locations}

In an accretion flow, the inflowing matter containing
multiple critical points experiences the discontinuous
transition of flow variables in the form of shock waves when the
Rankine-Hugoniot standing shock conditions are satisfied
\citep{Landau-Lifshitz59}. These conditions essentially
describe the state of the accretion flow on both sides of the shock
wave. At the shock front, these conditions are expressed as 
(a) Conservation of mass flux: $\dot{m}_{+} = \dot{m}_{-}$, 
(b) Conservation of energy flux: ${\cal E}_{+} = {\cal E}_{-}$
and 
(c) Conservation of momentum: 
$ W_{+} + \Sigma_{+}u^{2}_{+}=W_{-} + \Sigma_{-}u^{2}_{-}$,
where, all the quantities have their usual meaning and `$-$' and `$+$'
denote the quantities measured immediately before and after the shock 
transition. It is to be noted that in this study, we consider the shock to be 
thin, steady and non-dissipative in nature.

When the accretion flow possesses shock wave, inflowing matter experiences
compression while crossing the shock front. In a two temperature accretion 
flow, such compression arises primarily due to ions only because of their higher 
inertia compared to the electrons. Due to compression, as ions are slowed down,
kinetic energy of the ions is converted into thermal energy and electrons are
expected to get a part of this thermal energy via Coulomb coupling.
However, in reality, the relaxation time for electron-ion  
collision ($t_{\text{ei}}$) is generally  higher than both the electron-electron 
($t_{\text{ee}}$) and ion-ion ($t_{\text{ii}}$) collision time scales \citep{Colpi1984,Frank2002}.
Hence, electron temperature is expected to be mostly unaffected across the shock
front due to the shock transition. With this consideration, in this work, we assume  
$T_{e+} = T_{e-}$, immediately just before and after the shock.

In an accretion disc, sub-Keplerian flow 
starts it journey towards the black hole with almost negligible radial velocity 
after deviating from the Keplerian
disc \citep{Chakrabarti90}. Due to the gravitational
attraction, flow gains it radial velocity and becomes supersonic after crossing the
outer critical point ($x_{\rm out}$). Flow continues to move further before jumping to
the sub-sonic branch via shock transition and subsequently crosses the inner 
critical point ($x_{\rm in}$) before falling in to the black hole. In order to obtain such a 
global accretion solution including shock wave, for computational simplicity,
we begin our numerical integration of equations (18,21,22,23) from 
$x_{\rm out}$ and continue once outward up to $x_{\rm edge}$ for subsonic branch 
and then inward towards the black hole horizon for the supersonic branch. 
While calculating the supersonic branch, we search for the shock location
and associated inner critical point in the following way.  
Using a set of supersonic local flow variables, namely $x$, $u$, $a$, $M$ and
$\rho$, we calculate the 
total pressure and local energy of the flow.
Since these quantities along with the accretion rate, specific angular momentum and electron
temperature remain conserved across the shock front, we employ these conserved
quantities to calculate the same set of local flow variables for the subsonic branch
\citep{Chakrabarti2004}.
Employing these subsonic local flow variables, we continue integration 
towards the black hole horizon to find an inner critical point ($x_{\rm in}$) where $N$ and $D$
of equation (18)  tend to zero  simultaneously. Upon failing to obtain the inner
critical point, we choose another set of supersonic local flow variables by decreasing 
$x$ and again look for $x_{\rm in}$. We continue the process and once  $x_{\rm in}$
is found, the shock location ($x_s$) is obtained as $x_s=x$.
Needless to mention that the obtained
$x_{\rm in}$ represents an unique location corresponding to a given set of input
flow parameters including $x_{\rm out}$.  Using the inner critical point flow variables,
we again carry out the numerical integration further
and continue up to the inner edge of the disc.
With this, eventually we obtain a global accretion solution that passes through 
two saddle type critical point (namely, $x_{\rm in}$ and $x_{\rm out}$) and these points are
connected through the shock wave located at $x_s$. What is more is that following the above
procedure, we uniquely 
determine the shock location for a given set of input parameters chosen at the
outer critical point ($x_{\rm out}$). Conversely, one can repeat the same exercise
choosing the flow parameters at the inner critical point ($x_{\rm in}$) as well
while calculating the location of shock transition.

\begin{figure}
\includegraphics[width=0.485\textwidth]{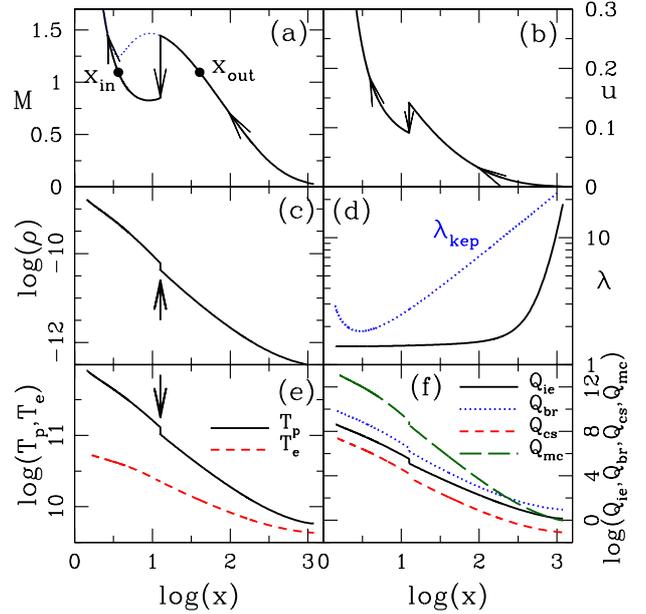}
\caption{Variation of (a) Mach number $M(=u/a)$, (b) radial velocity ($u$),
(c) density ($\rho$), (d) angular momentum ($\lambda$), (e) electron and ion 
temperature ($T_e, T_i$), and (f) cooling rates of the accretion flow as a 
function of radial distance. The shock solution presented here corresponds to 
result depicted by solid (black) curve in Fig. 2. See text for details.
}
\label{fig:fig_03}
\end{figure}

\subsection{Accretion solutions containing shock}

In Fig. 3a, we demonstrate the shock induced global accretion solution corresponding
to the result depicted using solid curve (black) in Fig. 2. 
Here, inflowing matter changes its sonic state after crossing the outer
critical point ($x_{\rm out}=40.594$) to become super-sonic and continues
to accrete towards the black hole. In principle, such supersonic matter
can smoothly enter into the black hole as indicated by the dotted curve (Fig. 3a).
However, we observe that the supersonic matter has the potential to join
with the subsonic branch passing through the inner critical point ($x_{\rm in}=3.6529$)
having $\lambda_{\rm in}=1.3947$ and $T_{\rm e, in}=4.074\times10^{10}{\rm K}$ 
through a shock transition.
We find that the shock conditions are favourable in both the
pre-shock and post-shock flow and these conditions are satisfied at
$x_{\rm s}=12.6370$, where the supersonic flow encounters a discontinuous
transitions of its variables to the subsonic branch. In the figure,
inner critical point ($x_{\rm in}$) and the outer critical point
($x_{\rm out}$) are marked and the solid vertical arrow indicates the shock jump. 
In Fig. 3b, we present the variation of radial velocity
corresponding the solution presented in Fig. 3a. Discontinuous transition
of radial velocity from supersonic to subsonic value is indicated by the
vertical arrow. As before, the direction of the arrows shows the 
direction of the flow motion towards the black hole. According to the second law of thermodynamics,
nature favors those accretion solutions that have high entropy content
\citep{Becker-Kazanas01}.
Incidentally, the supersonic inflowing matter passing through outer 
critical point possesses less entropy than the subsonic inflowing matter passing 
through inner critical point. 
Hence, supersonic inflowing matter prefers to follow the subsonic
branch of the solution while falling towards the black hole and this
happens through the discontinuous shock transition.
In Fig. 3c, we present the variation of density profile as function
of radial distance. We observe the compression of the inflowing matter in the 
post-shock region as density jumps sharply across the shock. This is shown 
by the solid curve where vertical arrow indicates density enhancement at shock. 
This happens mainly due to the reduction of radial velocity in the post-shock 
region.
In Fig. 3d, we show the radial variation of angular momentum for the same solution
as shown in Fig. 3a and it is denoted by the solid curve. We also depict
the Keplerian angular momentum distribution in the plot which is shown by the dotted
curve (blue).
The adopted viscosity prescription confirms the continuity of the angular momentum
distribution across the shock.
We continue our investigation of disc structure and in 
Fig. 3e, we show the temperature distribution of ion (obtained using sound speed) and electron
for the same solution as depicted in Fig. 3a, where solid and dashed curves represent
the results for ions and electrons, respectively. 
Here, the collisional
energy transfer between ions and electrons is largely inefficient and the
cooling time scale of the fast moving electrons are shorter than the slowly moving
ions. These considerations permit the inflowing matter to maintain two
temperature distribution. In addition to that ion temperature shoots up
discontinuously in the post shock region due to shock compression which is
denoted by the solid vertical arrow. The electron temperature remain unaffected
across the shock as we consider that the relaxation time for $e-i$ collision 
($t_{\text{ei}}$) is large compared to the $e-e$ and $i-i$ collision time scales 
\citep{Colpi1984,Frank2002}.
For instance, we calculate the ratio of time scales across the shock front which are
obtained as $t_{\rm ei}/t_{\rm ee} \sim 920$ and $t_{\rm ii}/t_{\rm ee} \sim 425$,
respectively. Fig.  3f shows the energy dissipation rates per
unit volume resulting from the Coulomb coupling, bremsstrahlung process, 
synchrotron process and inverse Comptonization of synchrotron photons,
respectively. Solid, dotted, dashed and long-dashed curves 
denote the radial variation of coulomb coupling $(Q_{ie})$, bremsstrahlung
cooling rate $(Q_{br})$, synchrotron cooling rate $(Q_{cs})$, and Comptonization
due to synchrotron photon $(Q_{mc})$. In the post-shock region all these cooling 
terms increases efficiently due to increase of density across the shock transition.

\begin{figure} 
\includegraphics[width=0.485\textwidth]{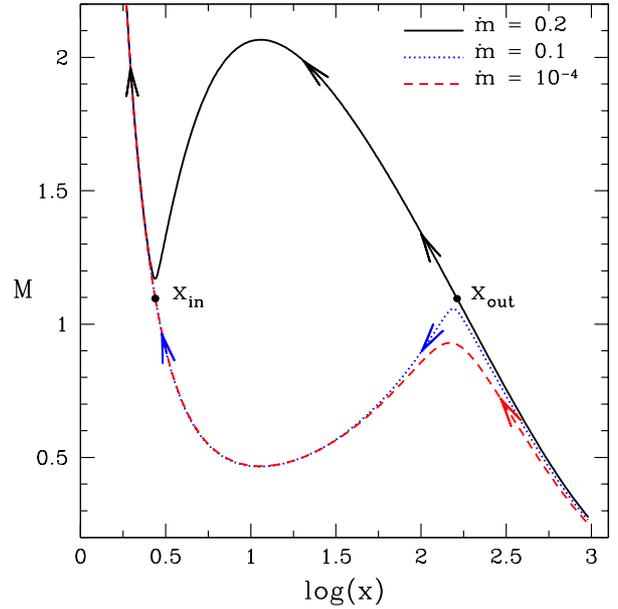}
\caption{Variation of Mach number $M(=u/a)$ with radial distance $x$ for 
accretion flows injected with various accretion rate (${\dot m}$) from the outer
edge of the disc.  Solutions plotted with dashed, dotted and solid curves are
for ${\dot m} =10^{-4}, 0.1$ and $0.2$, respectively. See text for details.}
\label{fig:fig_04}
\end{figure}

\begin{figure} 
\includegraphics[width=0.485\textwidth]{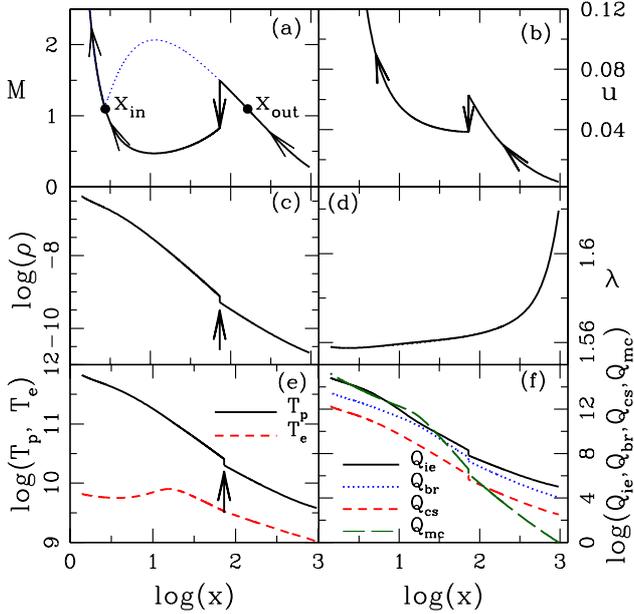}
\caption{Variation of (a) Mach number $M(=u/a)$, (b) radial velocity ($u$),
(c) density ($\rho$), (d) angular momentum ($\lambda$), (e) electron and ion 
temperature ($T_e, T_i$), and (f) cooling rates as a function of radial distance.
This shock solution corresponds to solid (black) curve in Fig. 4.
See text for details.}
\label{fig:fig_05}
\end{figure}

In \S 3.2, we have shown how the flow characteristics are altered when viscosity
is increased for flows having boundary conditions fixed at the outer edge of the disc.
In a similar way, here we continue our investigation while making an attempt to examine the role
of accretion rate in deciding the characteristic of the accretion flow. Towards
this, we choose
the outer boundary of the disc as $x_{\rm edge}=1000$, $\lambda_{\rm edge}=1.6257$, 
$T_{\rm e,~edge}=1\times10^9{\rm K}$, ${\cal E}_{\rm edge}=6.197\times10^{-4}$, ${\dot m}=10^{-4}$
and $\alpha=0.001$, respectively and obtain the global accretion solution passing through the 
inner critical point as $x_{\rm in}=2.7517$ before falling into the black hole. The result is depicted
in Fig. 4, where Mach number ($M$) is plotted as function of radial coordinate. In the figure, 
the dashed curve denotes the accretion solution corresponding to ${\dot m}=10^{-4}$. Next, we 
increase the accretion rate as ${\dot m}=0.1$ keeping the remaining flow parameters fixed
at $x_{\rm edge}$ and get the accretion solution analogous to the result
obtained for ${\dot m}=10^{-4}$. This result is depicted using dotted curve in Fig. 4.
When the accretion rate is increased 
further as ${\dot m}=0.2$, inflowing matter changes its character as it passes through 
the outer critical point ($x_{\rm out}=162.837$) instead of inner critical point and the result
is shown by the solid curve. After crossing $x_{\rm out}$, accretion flow 
becomes supersonic and continues to proceed towards the black hole. Eventually, 
centrifugal repulsion turns out to be comparable against gravity that renders the 
triggering of the discontinuous transition of flow variables in the form of shock wave.
In Fig. 5, we present the radial variation of Mach number ($M$), radial velocity 
($u$), density ($\rho$), angular momentum ($\lambda$), temperatures ($T_e, T_i$) and 
various cooling rates ($Q_{ie}, Q_{br}, Q_{cs}, Q_{mc}$) including shock transition for 
${\dot m}=0.2$. The general behavior of the above mentioned flow variables with radial 
coordinate are very much similar to the results presented in Fig. 3 and therefore, here we
skip their description to avoid repetition although Fig. 3f and Fig. 5f indicates 
that the contributions from the different cooling heads  are strongly depended on the
accretion rate which is indeed firmly expected.

\subsection{Shock Dynamics and Shock Properties}

\begin{figure} 
\includegraphics[width=0.485\textwidth]{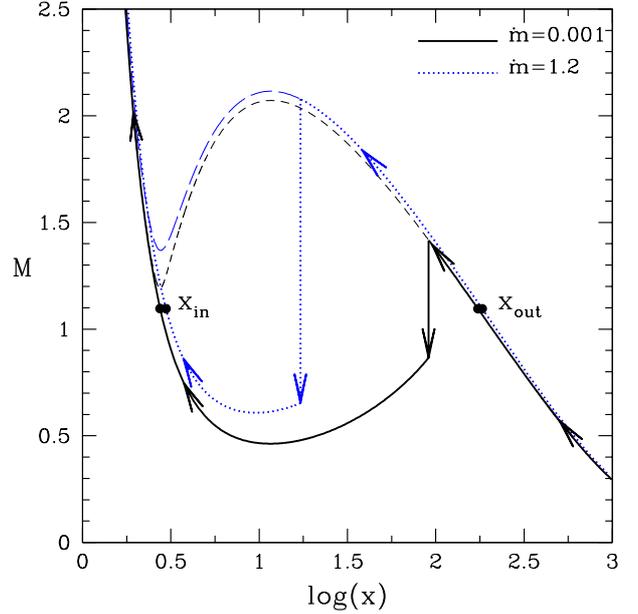}
\caption{Plot of global accretion solutions including shock waves. Flows with 
fixed outed boundary parameters are injected with different accretion rates
(${\dot m}$ which are marked in the figure. Each solution contains shock waves
which are shown by the vertical arrow as $x_s=91.107$ (solid) and $x_s = 16.757$ 
(dotted). See text for details.}
\label{fig:fig_06}
\end{figure}

In the course of shock study, it is customary to explore the influence of the 
global flow parameters, namely accretion rate (${\dot m}$) and viscosity ($\alpha$),
on the dynamical behavior of shock waves. Hence, as an 
example, we first choose an accretion flow having outer boundary parameters
as $x_{\rm edge}=1000$,  $\lambda_{\rm edge}=1.844$, 
$T_{\rm e,~edge}= 2\times10^9{\rm K}$, ${\cal E}_{\rm edge}=5.322\times10^{-4}$ 
and $\alpha = 0.005$, respectively and vary the accretion rate (${\dot m}$). 
When ${\dot m}= 0.001$ is considered, we find that flow encounters stationary 
shock transition at $x_s=91.107$. This result is depicted in Fig. 6, where the variation
of Mach number is plotted as function of radial coordinate. In the figure, solid curve denotes the
shock induced global accretion solution corresponding to ${\dot m}= 0.001$. 
Moreover, the solid vertical arrow indicates the location of shock transition 
and $x_{\rm in}$ and $x_{\rm out}$ denote the inner and outer critical points.
Overall, arrows indicate the direction of flow motion towards the black hole.
As ${\dot m}$ is increased, the effective radiative cooling efficiency in the 
post-shock region is enhanced that reduces the post-shock pressure. 
Eventually, shock front moves inward in order to maintain the pressure 
balance across the shock. Accordingly, we identify the maximum value of
the accretion rate as $\dot{m}^{\rm max}=1.2$ that allows the flow to pass 
through the stationary shock transition. Needless to mention that 
$\dot{m}^{\rm max}$ does not possess the universal value, instead it
strongly depends on the other flow parameters. For this accretion rate 
($\dot{m}^{\rm max}=1.2$), we calculate the global accretion solution
including shock and plot it using dotted curve in Fig. 6. As before,
vertical dotted arrow indicates the shock location at $x_s=16.787$.
When $\dot{m} > \dot{m}^{\rm max}$, accretion
solution including standing shock ceases to exist. However, non-steady shock
may still continue to present in the flow which could exhibit the time dependent 
physical processes similar to quasi-periodic oscillations (QPOs) as frequently 
observed in the case of Galactic black hole sources. Unfortunately,  the 
study of such scenarios are beyond the scope of the present paper.
We plan to consider the numerical investigation as a future project and
will be reported elsewhere.

\begin{figure} 
\includegraphics[width=0.475\textwidth]{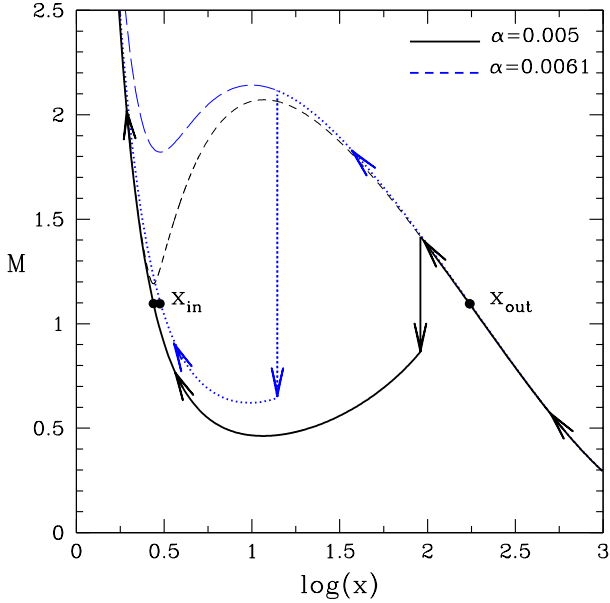}
\caption{
Same as Fig. 6, but the accretion solutions are obtained for different viscosity ($\alpha$)
which are marked in the figure. Each solution contains shock waves
which are shown by the vertical arrow as $x_s=91.107$ (solid) and $x_s = 13.969$ 
(dotted). See text for details. }
\label{fig:fig_07}
\end{figure}

In the subsequent analysis, we examine the role of viscosity ($\alpha$) in deciding 
the formation of stationary shock in an accretion flow having fixed outer boundary
conditions. As in the case of Fig. 6, here again we consider the accretion flow
having outer boundary conditions as $x_{\rm edge} = 1000$, $\lambda_{\rm edge}=1.844$,
$T_{\rm e,~edge}= 2\times10^9{\rm K}$, ${\cal E}_{\rm edge}=5.322\times10^{-4}$,
${\dot m}= 0.001$ and vary $\alpha$ to obtain the shock induced global accretion solutions.
When viscosity is chosen as $\alpha = 0.005$, we  find that flow encounters standing shock
transition at  $x_s=91.107$ which is denoted by the solid vertical arrow in Fig. 7.  Here again,
arrows indicate the direction of the flow motion during the course of accretion on
to the black hole. In addition, inner and outer critical points
are indicated in the figure as $x_{\rm in}$
and $x_{\rm out}$, respectively. When viscosity is increased, angular momentum of the
flow is being transported outwards more rapidly that causes the weakening of centrifugal 
repulsion against gravity. As a consequence, shock front settles down at a smaller radial
coordinate just to maintain the pressure balance across the shock. We calculate
the minimum shock location ($x_s^{\rm min}$) by increasing the $\alpha$ and find
$x_s^{\rm min}=13.969$ for $\alpha=0.0061$. Obviously, the obtained $x_s^{\rm min}$
depends on the other flow variables that we keep fixed here at the outer edge of the disc.
In the figure, we indicate the shock transition corresponding to $\alpha=0.0061$ by the dotted vertical 
arrow. When $\alpha>0.0061$, standing shock conditions are not favorable, and therefore,
stationary shock disappears. 

\begin{figure} 
\includegraphics[width=0.475\textwidth]{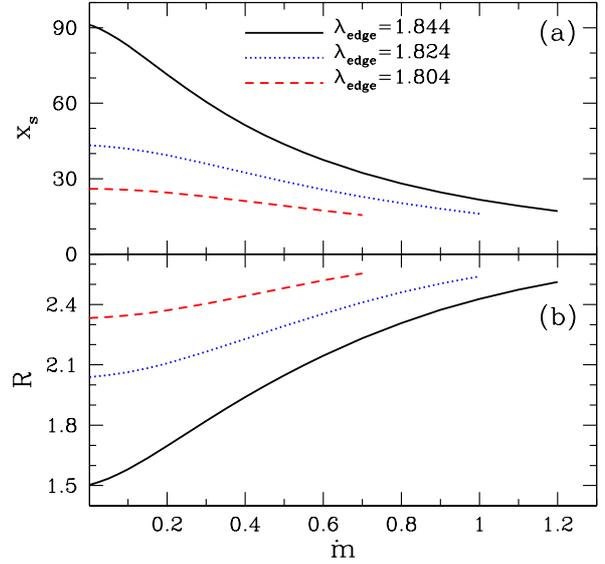}
\caption{Variation of (a) shock location ($x_{s}$) and (b) compression ratio ($R$)
with accretion rate $(\dot{m})$ for flows with fixed outer boundary parameters. 
Solid, dotted and dashed curves are for $1.844$, $1.824$ and $1.804$
respectively. See text for details.
}
\end{figure}

In Fig. 8, we investigate the shock properties as function of accretion rate ${\dot m}$.
Here, we choose the outer edge of the disc at $x_{\rm edge}=1000$ and inject
matter with ${\cal E}_{\rm edge}=5.322\times10^{-4}$, 
$T_{\rm e, edge}=2\times10^9{\rm K}$ and $\alpha =0.005$.
In the upper panel (Fig. 8a), we show the variation of shock location
corresponding to the different $\lambda_{\rm edge}$.
The solid, dotted and dashed curves are for $\lambda_{\rm edge}=1.844$, 
$1.824$ and $1.804$, respectively. For a given $\lambda_{\rm edge}$, 
as illustrated in the figure, shock front proceeds towards the black hole
horizon with the increase of accretion rate (${\dot m}$) irrespective to
the angular momentum at the outer edge. 
Evidently, it is not possible to get the shocked accretion solution when 
${\dot m}$ is exceedingly large.
This is because there exists a maximum value of accretion rate (${\dot m^{\rm max}}$ )
corresponding to the chosen outer boundary flow parameters. When 
${\dot m} > {\dot m^{\rm max}}$, the possibility of shock formation fails
as shock conditions are not satisfied there. Moreover,
for a given ${\dot m}$ we find the shock to form further out
for flows with higher $\lambda_{\rm edge}$. This establishes the
fact that the possibility of shock transition is perhaps centrifugally driven.
Meanwhile, we have pointed out in the Fig. (3) and Fig. (5) that density is
enhanced in the post-shock region due to shock compression. Since the 
radiative cooling processes largely depend on the matter density, emergent 
radiations from the disc are also rely upon it and therefore, it is worthy 
to examine the density jump of the flow across the shock.
Accordingly, we calculate the density
compression of the flow across the shock by computing the compression
ratio which is defined as the ratio of vertically averaged post-shock 
and the pre-shock density ($R=\Sigma_{+}/\Sigma_{-}$). 
In Fig. 8b, we present
the variation of the compression ratio ($R$) with ${\dot m}$ for the same
set of flow parameters as used in Fig. 8a. With the increase of ${\dot m}$,
as the shock front proceeds towards the black hole horizon, post-shock flow 
experiences further compression and hence, $R$ increases with the
increase of ${\dot m}$.

\begin{figure} 
\includegraphics[width=0.485\textwidth]{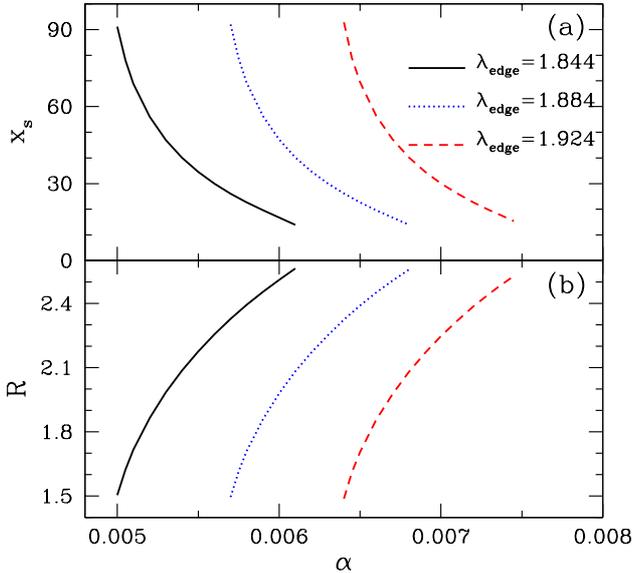}
\caption{Variation of (a) shock location ($x_{s}$) and (b) compression ratio ($R$)
as function of viscosity ($\alpha$) for flows injected with fixed outer boundary
parameters. Solid, dotted and dashed curves are for
$\lambda_{\rm edge} = 1.844, 1.884$ and $1.924$, respectively. See text for details.
}
\end{figure}

In the next, we carry out the investigation of shock properties by varying
viscosity $\alpha$ for flows with fixed outer boundary conditions.
The obtained results are presented in Fig. 9 where we choose $x_{\rm edge}=1000$,
${\cal E}_{\rm edge}=5.322\times10^{-4}$, $T_{\rm e, edge}=2\times10^9{\rm K}$ 
and ${\dot m}=0.001$, respectively. Here, solid, dotted and dashed curves 
represented the results corresponding to $\lambda_{\rm edge}=1.844$, 
$1.884$ and $1.924$, respectively.  As illustrated in Fig. 9a, shock front 
proceeds towards the black hole horizon with the increase of $\alpha$ irrespective to
the angular momentum at the outer edge. For a given $\lambda_{\rm edge}$, when $\alpha$
is increased, $\lambda(x)$ is decreased along the flow motion. This weakens the
centrifugal repulsion against the gravity and causes the shock front to move
inward. Moreover, for a fixed $\alpha$, shock forms at a larger distance from
the horizon when $\lambda_{\rm edge}$ is increased. This happens because
large $\lambda_{\rm edge}$ increases the strength of the centrifugal barrier
and therefore, the shock front is pushed outside. 
This findings again clearly indicates that
the centrifugal force eventually plays the pivotal role in deciding the
formation of shock wave in an accretion disc. In addition, 
there exists a limiting 
range of $\alpha$ associated to a given outer boundary conditions of the flow. 
Out side this range of $\alpha$, standing shock conditions are not favorable and
therefore, global accretion solution including shock wave ceases to exist.
As before, we calculate the compression ratio ($R$) across the shock
front corresponding to the shocked accretion solution depicted in
Fig. 9a. We plot the obtained results in Fig. 9b where we observe positive 
correlation between $R$ and $\alpha$.

\section{Shock Parameter Space}

\begin{figure} 
\includegraphics[width=0.475\textwidth]{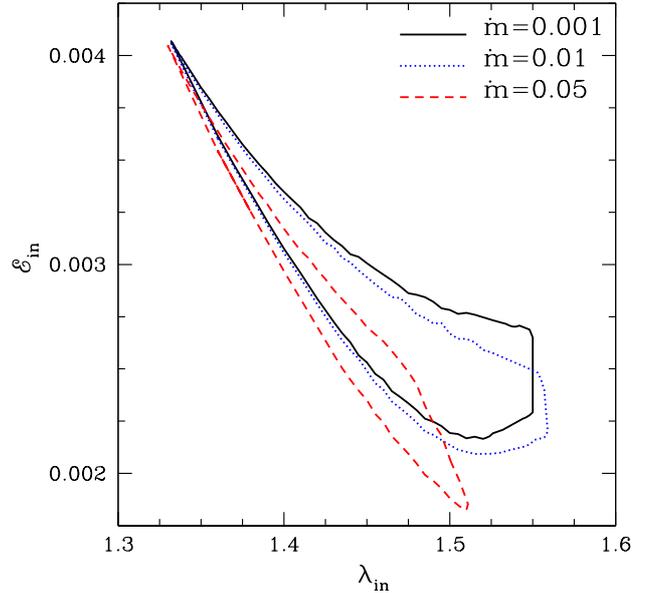}
\caption{Plot of $\lambda_{\rm in}-{\cal E}_{\rm in}$ parameter space for shock. 
Region bounded by solid, dotted and dashed curves for ${\dot m}=0.001, 0.01$ and
$0.05$, respectively. As ${\dot m}$ is enhanced, parameter space is shifted to
lower energy side due to energy dissipation in the form of radiative cooling.
See text for details.
}
\label{fig:fig_10}
\end{figure}

It is already observed that the two-temperature transonic accretion flows 
around black holes possess shock waves and accretion solutions of this kind 
are not the discrete solutions, in fact they exist for a wide range of flow 
parameters. To validate this claim and also to 
understand the influence of flow parameters on the
shock induced global accretion solutions, we separate the regions of
parameter space spanned by the
angular momentum ($\lambda_{\rm in}$, measured at $x_{\rm in}$)  
and the local energy of the flow (${\cal E}_{\rm in}$, measured at $x_{\rm in}$) 
that allows shocked accretion solutions.  
We calculate the shock parameter space at $x_{\rm in}$ 
due to the fact that the acceptable range of the angular
momentum at the inner edge of the disc and the location of the inner 
critical points are $1.2 \lesssim \lambda_{\rm in} \lesssim 2$ and
$2 \lesssim x_{\rm in} \lesssim 4$, respectively \citep{Chakrabarti90,Chakrabarti2004}. 
Therefore, we can safely assume that the flow enters in to the black hole with angular
momentum and energy similar to their values measured at $x_{\rm in}$. 
In Fig. 10, we present the classification of shock parameter space in terms of the
accretion rate (${\dot m}$). The region bounded by the solid, dotted and dashed curves 
are obtained for ${\dot m}=0.001, 0.01$ and $0.05$, respectively.
Here, we choose $\alpha =0.01$ and $T_{\rm e, in}=6.75\times10^9{\rm K}$.
As ${\dot m}$ is increased, the effective region of the parameter space for 
shock is shrunk. When the adopted cooling processes are active in the flow,
inflowing matter loses its energy while accreting towards the black hole.
This eventually causes the parameter space to shift in the lower energy  
domain for increasing accretion rate of the flow. Moreover, 
when ${\dot m}$ gradually increases, beyond its critical limit, the parameter 
space disappears completely as standing shock conditions fail to satisfy there.

\begin{figure} 
\includegraphics[width=0.475\textwidth]{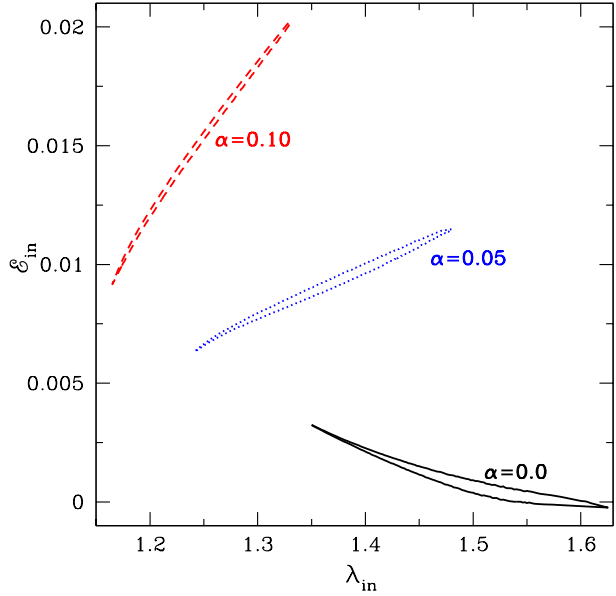}
\caption{Plot of $\lambda_{\rm in}-{\cal E}_{\rm in}$ parameter space for shock. 
Region bounded by different line styles are for different viscosity parameter ($\alpha$).
The lowermost parameter space is drawn for $\alpha=0$. As $\alpha$ is increased
($\Delta \alpha = 0.05$) which are marked in the plot, shock parameter space is 
shifted to higher energy and lower angular momentum domain. See text for details.
}
\label{fig:fig_11}
\end{figure}

We continue our investigation to study the variation of the shock parameter 
space for flows with varying viscosity parameters ($\alpha$) and present the
results in Fig. 11, where the shock parameter space corresponding to a given 
viscosity parameter is marked. Here, we choose ${\dot m}=0.05$ and 
$T_{\rm e, in}=6.75\times10^9{\rm K}$, respectively. We observe that shock
parameter space is shifted towards the lower angular momentum and higher
energy domain when $\alpha$ is increased.
This is the consequence of the dual effects of viscosity in an accretion flow
where viscosity not only transports the angular momentum outwards, but also
dissipates energy as well. Moreover, we observed that standing shocks can 
form even for a low angular momentum flows provided the viscosity is chosen 
to be sufficiently large. However, it is not possible to increase the viscosity indefinitely.
This is because above a critical viscosity limit, standing shock transition in
an accretion flow does not occur and consequently, shock parameter space
disappears. It is to be noted that the value of the critical viscosity is not
universal, instead it largely depends on the other parameters of the flow.

\begin{figure} 
\includegraphics[width=0.475\textwidth]{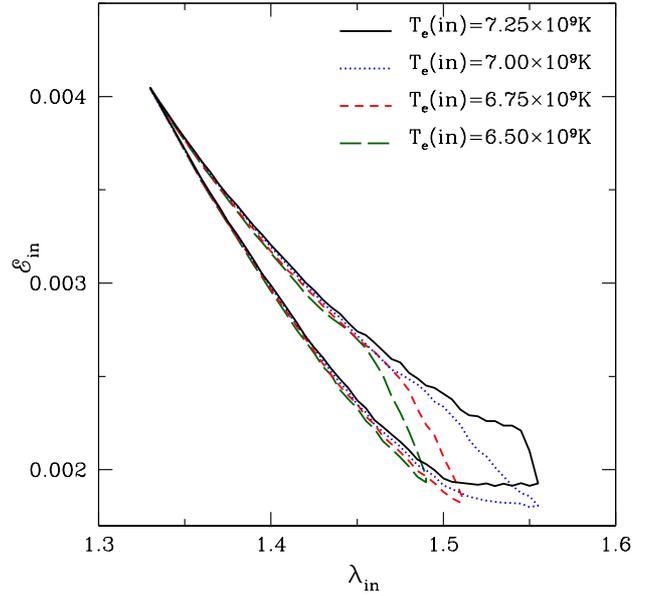}
\caption{Modification of $\lambda_{\rm in}-{\cal E}_{\rm in}$ shock parameter 
space as function of  $T_{\rm e,in}$. $T_{\rm e,in}$ values are marked in the figure
that denote parameter spaces separated with different line styles. See text for
details.
}
\label{fig:fig_12}
\end{figure}

We further classify the shock parameter space as function of $T_{\rm e, in}$ and
present the results in Fig. 12. Here, we choose ${\dot m}=0.05$ and
$\alpha=0.01$, respectively. The shock parameter spaces are separated using
boundaries drawn with long-dashed, dashed, dotted and solid curves which
are obtained for $T_{\rm e, in}=6.50\times10^9{\rm K} , 6.75\times10^9{\rm K},
7.00 \times10^9{\rm K}$ and $7.25\times10^9{\rm K}$, 
respectively. We find that the effective region of the shock parameter space is 
reduced with the decrease of $T_{\rm e, in}$. This is not surprising because in 
order to obtain an accretion solution having smaller  $T_{\rm e, in}$, the overall
cooling efficiency needs to be higher and the increasing cooling effect reduces the
effective region of the parameter space for standing shock (see Fig. 10). In addition,
Fig. 12 clearly indicates that there exists a range of $T_{\rm e, in}$ which 
allows standing shock transition for a given set of input parameters of the flow.

\begin{figure} 
\includegraphics[width=0.475\textwidth]{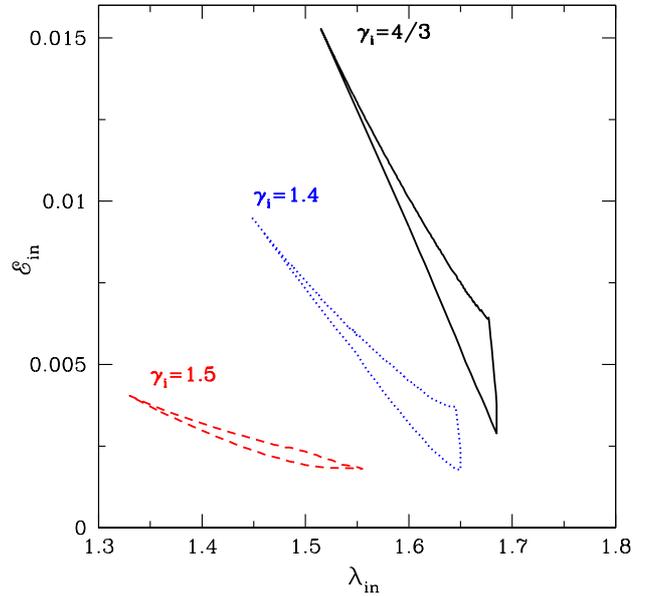}
\caption{Modification of parameter space for standing shock as function of
ions specific heats ratio ($\gamma_i$). Parameter spaces bounded with 
solid, dotted and dashed curves are for $\gamma_i = 4/3, 1.4$ and $1.5$,
respectively. See text for details.
}
\label{fig:fig_13}
\end{figure}

In a two-temperature accretion flow, since electrons are lighter than the protons,
it is customary to treat electrons as relativistic in nature while protons are considered 
as non-relativistic. Accordingly, as a conservative estimate, so far we have chosen the adiabatic 
indices  for electrons and ions as $\gamma_{e}=4/3$ and $\gamma_{i}=3/2$,
respectively. However, in a realistic scenario, $\gamma_{i}$ is expected to vary 
depending on the ratio between the thermal energy and the rest mass energy 
of the ions \citep{Frank2002}. In order to understand the effect 
of $\gamma_{i}$ on the two-temperature global shock solutions, we calculate
the parameter space for standing shock similar to Fig. 10 for three different values
of $\gamma_{i}$ and depict the result in Fig. 13, where the parameter spaces separated
with solid, dotted and dashed boundary are for $\gamma_i = 4/3, 1.4$ and $1.5$,
respectively. Here, we choose $\alpha =0.01$, ${\dot m}=0.05$ and $T_{\rm e, in}=7\times10^9{\rm K}$.
We find that standing shock exists in all three cases and also observe that 
the effective region of the parameter space for standing shock is large for
lower values of $\gamma_{i}$ and it reduces considerably when ions changes
its character towards the non-relativistic limit.

\section{Conclusions} 

In this paper, we provide a self-consistent formalism to study the properties of
a two-temperature accreting flow around a Schwarzschild black hole. We
choose the inflowing matter to be rotating, viscous and advection dominated which 
is confined within a thin disc structure. Radiative processes, namely, bremsstrahlung, 
cyclo-synchrotron and Comptonization of synchrotron photons processes are
considered to be active in the flow. Moreover, we consider that the interaction 
between electrons and ions is guided by the Coulomb coupling process. During
the course of accretion towards the black hole,
rotating matter experiences centrifugal repulsion that eventually triggers the
discontinuous transition of flow variables in the form of shock waves.
Based on the consideration of second law of thermodynamics,
shock induced two-temperature global accretion solutions around 
black holes are thermodynamically preferred over the smooth solutions as the
former possess high entropy content 
\citep{Becker-Kazanas01}. With this, for the first time to our knowledge, we obtain
the self-consistent two-temperature global accretion solutions including shock waves
by solving the coupled hydrodynamic equations for ions and electrons simultaneously. 
These shocked accretion solutions seem to be essential to understand the observed spectral 
and timing properties of the black hole candidates \citep{Chakrabarti_Manickam00,
Nandi_etal01a,Nandi_etal01b,Nandi_etal12,Radhika-Nandi14,Iyer-etal15,Radhika-etal16a,
Radhika-etal16b}.

We obtain the two-temperature global accretion
solutions containing shock waves in presence of dissipation processes, namely, 
radiative coolings and viscosity (Fig. 3 and Fig. 5). We observe that due to the 
discontinuous shock transition, post-shock flow (equivalently PSC) is compressed
that causes the PSC to become hot and dense. Incidentally, synchrotron and
bremsstrahlung photons in PSC are likely to be reprocessed 
via inverse Comptonozation to produce hard radiations 
\citep{Chakrabarti_Titarchuk95,Mandal2005} which are commonly
observed from the accretion systems harboring  black hole candidates. This
clearly gives a hint that the size of the PSC ($i. e.$, shock location, $x_s$) perhaps 
play key role in emitting the high energy photons from the disc. Moreover, for an accretion flow 
injected from a fixed outer edge, we find that the dynamics of the shock front is regulated by
the dissipation parameters. When accretion rate (${\dot m}$) is increased, the effect of cooling
becomes more efficient in the post-shock flow resulting the reduction of 
thermal pressure at PSC. In order to maintain the balance of total pressure across
the shock front, PSC is forced to settle down to a smaller radius
from the black hole (Fig. 6). Similarly, as the viscosity ($\alpha$) is increased, 
the efficiency of angular momentum transport  
in the outward direction is enhanced causing
the weakening of centrifugal repulsion against gravity. This eventually
compels the shock front to move towards the black hole just to maintain 
pressure balance across the shock front (Fig. 7). Overall, in our two-temperature 
model, we find that shock induced global accretion solutions are not isolated
solutions as shock exists for a wide range of dissipation parameters.
Interestingly, when dissipation parameters exceed there critical limits
for flows with fixed injection boundary, PSC disappears as the standing
shock conditions fail to satisfy for a highly dissipative flow (Fig. 8-9).

One of the motivation of this work is to identify the ranges of flow parameters
that allow shock in two-temperature accretion flow. In Fig. 10, we explore the
parameter space within which the two-temperature accretion flow with shock can
exist as function of angular momentum and energy both measured at the 
inner critical point
for various values of  accretion rate (${\dot m}$). As ${\dot m}$ is increased,
dissipation due to cooling is increased and flow loses its energy while advecting 
towards the black hole. This reduces the possibility of shock formation and effectively
the shock parameter space is shrunk and shifted towards the lower energy side.
Further, we separate the parameter space for shock as function of viscosity ($\alpha$) 
as well and find that shock region is reduced and displaced towards the higher
energy and lower angular momentum domain as $\alpha$ is increased (Fig. 11). 
This happens due to the dual role of viscosity as the increase of $\alpha$ 
enhances the outward diffusion of angular momentum and also dissipates
energy in the flow. Moreover, we calculate the shock parameter space as function
of electron temperature at the inner critical point ($T_{\rm e, in}$) and find the
the effective region of the shock parameter space is reduced with the decrease 
of $T_{\rm e, in}$ (Fig.12). 

Since ions are more massive than electrons, in this work, we treat ions as 
thermally non-relativistic  and consider the adiabatic index for ions in most of
cases as $\gamma_{i}=3/2$. However, in reality, the theoretical limit of
$\gamma_{i}$ lies in the range between $4/3$ to $5/3$ depending on the
ratio between the thermal energy and the rest mass energy of the ions
\citep{Frank2002}. To infer this, we calculate the shock parameter space for 
three different $\gamma_{i}$ and observe that global shock solutions
exist in all cases. Moreover, we find that shock parameter space is 
shrunk as $\gamma_{i}$ is increased which essentially indicates that
possibility of shock formation is reduced for higher $\gamma_{i}$.

The self-consistent two-temperature global shocked accretion solutions have
extensive implications in the context of modeling of observational features of
black hole sources. In fact, these hydrodynamical shock solutions in presence 
of the relevant cooling mechanisms are perhaps essential to calculate the 
emission spectrum of a black hole system \citep{Chakrabarti_Titarchuk95,
Mandal2005}. So far, several theoretical attempts were made to examine
the spectral behavior of black hole sources considering shock properties in a 
parametric way \citet{Mandal2005,ChakrabartiMandal2006}. In order for generalization, 
a study of the spectral properties of  black holes involving the self-consistent 
two-temperature shocked accretion solution is under progress  and will be 
reported elsewhere.

Finally, we  discuss the limitations of the present work as it is developed based
on some approximations. We use pseudo-Newtonian gravitational potential 
to describe the space-time geometry around the black hole which allows us 
to study the non-linear shock phenomenon and its properties avoiding the
complicated general relativistic calculations. Moreover, we consider the constant 
adiabatic indices for both ions and electrons instead of estimating them 
self-consistently based on the thermal properties of the flow
\citep{Ryu2006,Kumar2013,Kumar2014} . 
Further, we choose stochastic
magnetic field configuration rather ignoring the large scale magnetic
fields and also neglect the effects of radiation pressure. Of course, 
the implementation of such issues is beyond the scope of present paper.
However, we believe that the basic conclusions of this work will remain 
qualitatively unaltered due to the above approximations.

\section*{Acknowledgments}
We thank the reviewer for useful suggestions and comments that help us to
improve the manuscript. 

\section*{Appendix: Calculation of radial velocity gradient at critical point}

The radial velocity gradient at the critical point ($x_c$) has the form $(du/dx)=0/0$. 
Hence, in order to understand the behavior of the flow properly, we apply the 
l$^{\prime}$Hospital rule at $x_c$ \citep{Kopp_Holzer_1976,Holzer_1977,
Jacques_1978} and therefore, obtain the expression of the radial velocity gradient
at the critical point ($x_c$) as, 

$$
\left( \frac{du}{dx}\right)_c=\frac{N_1-D_2\pm\sqrt{(N_1-D_2)^2+4D_1N_2}}{2D_1},
\eqno(A1)
$$
where
$$\begin{aligned}
N_1= &\Lambda_{11}-\Gamma_{21}-\frac{\frac{\lambda^2}{x^3}-F}{\gamma_i-1}
-\frac{\gamma_ia}{\gamma_i-1}-\frac{\gamma_ia^2}{\gamma_i-1}\left(\frac{3}{2x}
-\frac{1}{2F}\frac{dF}{dx}\right)\\
&-\frac{u}{\gamma_i-1}\frac{2\lambda\lambda_{11}}{x^3}
-\frac{\gamma_iau}{\gamma_i-1}a_{121}\\
&-\bigg[\frac{\gamma_iu}{\gamma_i-1}+\frac{2\gamma_iau}{\gamma_i-1}
\left(\frac{3}{2x}-\frac{1}{2F}\frac{dF}{dx}\right)\bigg]a_{11},\\
N_2=&\Lambda_{12}-\Gamma_{22} -\frac{u}{\gamma_i-1}
\left(\frac{2\lambda\lambda_{12}}{x^3}-\frac{3\lambda^2}{x^4} 
- \frac{dF}{dx}\right)-\frac{\gamma_iau}{\gamma_i-1}a_{122}\\
&-\frac{\gamma_ia^2u}{\gamma_i-1}\left(-\frac{3}{2x^2}
+\frac{1}{2F^2}\left(\frac{dF}{dx}\right)^2-\frac{1}{2F}\frac{d^2F}{dx^2}\right)\\
&-\bigg[\frac{\gamma_iu}{\gamma_i-1}+\frac{2\gamma_iau}{\gamma_i-1}
\left(\frac{3}{2x}-\frac{1}{2F}\frac{dF}{dx}\right)\bigg]a_{12},\\
D_1=&\frac{4\gamma_ia}{\gamma_i-1}a_{11}-2\frac{\gamma_i+1}{\gamma_i-1}u+\Gamma_{11},\\
D_2=&\frac{4\gamma_ia}{\gamma_i-1}a_{12} + \Gamma_{12},\\
\Lambda_{11}=&-\left(\Lambda_{ie}+\Lambda_{ib}\right)
\left( \frac{a_{11}}{a} + \frac{1}{u}\right) \\
&+ \left(\frac{\Lambda_{ie}}{T_i-T_e}  + \frac{\Lambda_{ib}}{2T_i}\right)T_{21} 
- \Lambda_{ie}\left(\frac{1}{T_i-T_e}  - \frac{3}{2T_e}\right)T_{11},\\
\Lambda_{12}=&-(\Lambda_{ie}+\Lambda_{ib})\left(\frac{a_{12}}{a} + \frac{3}{2x} 
- \frac{1}{2F}\frac{dF}{dx}\right)\\
&+ \left(\frac{\Lambda_{ie}}{T_i-T_e} + \frac{\Lambda_{ib}}{2T_i}\right)T_{22} 
- \Lambda_{ie}\left(\frac{1}{T_i-T_e} - \frac{3}{2T_e}\right)T_{12},\\
\Gamma_{11}=& \frac{-4\alpha I_n(gaa_{11} + u)\lambda_{11}}{x} - 
  \frac{2\alpha I_n(ga^2+u^2)\lambda_{111}}{x},\\
\end{aligned}$$
$$\begin{aligned}
 \Gamma_{12}=&\frac{-4\alpha I_n(gaa_{12})\lambda_{11}}{x} 
- \frac{2\alpha I_n(ga^2+u^2)\lambda_{112}}{x} \\
& + \frac{2\alpha I_n(ga^2+u^2)\lambda_{11}}{x^2},\\
 \Gamma_{21}=& \frac{-4\alpha I_n(gaa_{11} + u)\lambda_{12}}{x} - 
  \frac{2\alpha I_n(ga^2+u^2)\lambda_{121}}{x} \\
&+ \frac{8\alpha I_n(gaa_{11} + u)\lambda}{x^2}
  + \frac{4\alpha I_n(ga^2+u^2)\lambda_{11}}{x^2},\\
\Gamma_{22}=&\frac{-4\alpha I_n g a a_{12}\lambda_{12}}{x} - \frac{2\alpha I_n(ga^2+u^2) \lambda_{122}}{x} \\
& + \frac{2\alpha I_n(ga^2+u^2)\lambda_{12}}{x^2} 
  + \frac{8\alpha I_n gaa_{12}\lambda}{x^2} \\
 & + \frac{4\alpha I_n(ga^2+u^2)\lambda_{12}}{x^2} - \frac{8\alpha I_n(ga^2+u^2)\lambda}{x^3},\\
a_{111}=&\left(\frac{1}{u} + \frac{u}{a^2}\right)a_{11} - \left(\frac{1}{a} 
   + \frac{a}{u^2}\right),\\
a_{112}=&\left(\frac{1}{u} + \frac{u}{a^2}\right)a_{12},\\
a_{121}=&-\frac{1}{a^2}\bigg\{\frac{\lambda^2}{x^3} - F\bigg\}a_{11} 
 + \bigg\{\frac{3}{2x} + \frac{1}{2F}\frac{dF}{dx}\bigg\}a_{11} 
 + \frac{2\lambda\lambda_{11}}{ax^3},\\
a_{122}=&-\frac{1}{a^2}\bigg\{\frac{\lambda^2}{x^3} - F\bigg\}a_{12}
 + \bigg\{\frac{3}{2x} + \frac{1}{2F}\frac{dF}{dx}\bigg\}a_{12} 
 + \frac{2\lambda\lambda_{12}}{ax^3} \\
 &- \frac{1}{a}\left(\frac{3\lambda^2}{x^4}
 + \frac{dF}{dx}\right) + a\bigg\{-\frac{3}{2x^2}+\frac{1}{2F^2}
 \left(\frac{dF}{dx}\right)^2-\frac{1}{2F}\frac{d^2F}{dx^2}\bigg\},\\
 \lambda_{111}=&\frac{\alpha(ga^2 - u^2)}{u^3} - \frac{2\alpha(gaa_{11} - u)}{u^2} \\
& + \frac{2\alpha gaxa_{11}}{u}\left(\frac{a_{11}}{a}  -\frac{1}{u} + \frac{a_{111}}{a_{11}}\right),\\
\lambda_{112}=&- \frac{2\alpha gaa_{12}}{u^2} +\frac{2\alpha gaxa_{11}}{u}
 \left(\frac{1}{x} + \frac{a_{12}}{a} + \frac{a_{112}}{a_{11}}\right),\\
\lambda_{121}=& \frac{2\alpha gaxa_{12}}{u}\left(\frac{a_{11}}{a} -\frac{1}{u} + \frac{a_{121}}{a_{12}}\right)\\
&- \frac{\alpha(ga^2+u^2)}{u^2}  + \frac{2\alpha(gaa_{11} + u)}{u},\\
\lambda_{122}=& \frac{2\alpha gaxa_{12}}{u}\left(\frac{1}{x} + \frac{a_{12}}{a} 
 +\frac{a_{122}}{a_{12}}   \right)+ \frac{2\alpha gaa_{12}}{u},\\ 
 a_{11}=&\frac{a}{u} - \frac{u}{a},\\
a_{12}=&\frac{1}{a}\bigg\{\frac{\lambda^2}{x^3} - F\bigg\}
 + \bigg\{\frac{3a}{2x} + \frac{a}{2F}\frac{dF}{dx}\bigg\},\\
\lambda_{11}=&- \frac{\alpha(ga^2 - u^2)}{u^2} + \frac{2\alpha gaxa_{11}}{u},\\
\lambda_{12}=&\frac{2\alpha gaxa_{12}}{u}+\frac{\alpha(ga^2 + u^2)}{u},\\
\end{aligned}$$
$$\begin{aligned}
T_{11}=&-\frac{(\gamma_e - 1)a_{11}T_e}{a} - \frac{(\gamma_e - 1)T_e}{u},\\
T_{12}=& \frac{(\gamma_e -1)m_p\mu_e}{k_{B}u}(\Lambda_e - \Gamma_e)\\
& - (\gamma_e - 1)T_e\bigg\{\frac{3}{2x} + \frac{1}{2F(x)}\frac{dF(x)}{dx}\bigg\}
 -\frac{(\gamma_e - 1)a_{12}T_e}{a},\\
T_{21}=&\left(\frac{2m_p\mu_paa_{11}}{k_B}-\frac{\mu_p}{\mu_e}T_{11}\right) {\rm and} \\
T_{22}=&\left(\frac{2m_p\mu_paa_{12}}{k_B}-\frac{\mu_p}{\mu_e}T_{12}\right).
\end{aligned}$$

\noindent Here, all the quantities have their meaning already defined earlier in the text.



\begin{thebibliography}{68}
\expandafter\ifx\csname natexlab\endcsname\relax\def\natexlab#1{#1}\fi

\bibitem[{{Abramowicz} {et~al}\mbox{.}(1995){Abramowicz}, {Chen}, {Kato},
  {Lasota}, \& {Regev}}]{abramowicz1994thermal}
{Abramowicz} M.~A., {Chen} X., {Kato} S., {Lasota} J.-P., {Regev} O., 1995,
  \apjl, 438, L37

\bibitem[{{Abramowicz} {et~al}\mbox{.}(1988){Abramowicz}, {Czerny}, {Lasota},
  \& {Szuszkiewicz}}]{abramowicz1988}
{Abramowicz} M.~A., {Czerny} B., {Lasota} J.~P., {Szuszkiewicz} E., 1988, \apj,
  332, 646

\bibitem[{{Aktar}, {Das} \& {Nandi}(2015){Aktar}, {Das}, \&
  {Nandi}}]{aktar2015}
{Aktar} R., {Das} S., {Nandi} A., 2015, \mnras, 453, 3414

\bibitem[\protect\citeauthoryear{Aktar et al.}{2017}]{Aktar-etal17} 
Aktar R., Das S., Nandi A., Sreehari H., 2017, MNRAS, 471, 4806

\bibitem[{{Becker} \& {Kazanas}(2001)}]{Becker-Kazanas01}
{Becker} P.~A., {Kazanas} D., 2001, \apj, 546, 429

\bibitem[{{Chakrabarti}(1989)}]{Chakrabarti1989}
{Chakrabarti} S.~K., 1989, \apj, 347, 365

\bibitem[\protect\citeauthoryear{Chakrabarti}{1990}]{Chakrabarti90} 
Chakrabarti S.~K., 1990, {Theory of Transonic Astrophysical Flows}

\bibitem[{{Chakrabarti} \& {Titarchuk}(1995)}]{Chakrabarti_Titarchuk95}
{Chakrabarti} S., {Titarchuk} L.~G., 1995, \apj, 455, 623

\bibitem[{{Chakrabarti} \& {Molteni}(1995)}]{Chakrabarti_Molteni1995}
{Chakrabarti} S.~K., {Molteni} D., 1995, \mnras, 272, 80

\bibitem[{{Chakrabarti}(1996)}]{Chakrabarti96}
{Chakrabarti} S.~K., 1996, \apj, 464, 664

\bibitem[{{Chakrabarti}(1999)}]{Chakrabarti99}
{Chakrabarti} S.~K., 1999, \aap, 351, 185

\bibitem[{{Chakrabarti} \& {Manickam}(2000)}]{Chakrabarti_Manickam00}
{Chakrabarti} S.~K., {Manickam} S.~G., 2000, \apjl, 531, L41

\bibitem[{{Chakrabarti} \& {Das}(2004)}]{Chakrabarti2004}
{Chakrabarti} S.~K., {Das} S., 2004, \mnras, 349, 649

\bibitem[{{Chakrabarti} \& {Mandal}(2006)}]{ChakrabartiMandal2006}
{Chakrabarti} S.~K., {Mandal} S., 2006, \apjl, 642, L49

\bibitem[{{Chen}(1995)}]{chen1995hot}
{Chen} X., 1995, \mnras, 275, 641

\bibitem[{{Chen} \& {Taam}(1995)}]{chen1994variability}
{Chen} X., {Taam} R.~E., 1995, \apj, 441, 354

\bibitem[{{Colpi}, {Maraschi} \& {Treves}(1984){Colpi}, {Maraschi}, \&
  {Treves}}]{Colpi1984}
{Colpi} M., {Maraschi} L., {Treves} A., 1984, \apj, 280, 319

\bibitem[{{Das}, {Chattopadhyay} \& {Chakrabarti}(2001){Das}, {Chattopadhyay},
  \& {Chakrabarti}}]{Das2001a}
{Das} S., {Chattopadhyay} I., {Chakrabarti} S.~K., 2001, \apj, 557, 983

\bibitem[{{Das} {et~al}\mbox{.}(2001){Das}, {Chattopadhyay}, {Nandi}, \&
  {Chakrabarti}}]{Das2001b}
{Das} S., {Chattopadhyay} I., {Nandi} A., {Chakrabarti} S.~K., 2001, \aap, 379,
  683

\bibitem[{{Das}(2007)}]{Das2007}
{Das} S., 2007, \mnras, 376, 1659

\bibitem[{{Das} \& {Chattopadhyay}(2008)}]{Das_Chattopadhyay08}
{Das} S., {Chattopadhyay} I., 2008, \na, 13, 549

\bibitem[{{Das} {et~al}\mbox{.}(2014){Das}, {Chattopadhyay}, {Nandi}, \&
  {Molteni}}]{Das2014}
{Das} S., {Chattopadhyay} I., {Nandi} A., {Molteni} D., 2014, \mnras, 442, 251

\bibitem[{{Dihingia}, {Das} \& {Mandal}(2015){Dihingia}, {Das}, \&
  {Mandal}}]{dihingia2015shocks}
{Dihingia} I.~K., {Das} S., {Mandal} S., 2015, in Astronomical Society of India
  Conference Series, Vol.~12, Astronomical Society of India Conference Series

\bibitem[{{Ding} {et~al}\mbox{.}(2000){Ding}, {Lan-Tian}, {Xue-Bing}, \&
  {Ye}}]{shi2000radial}
{Ding} S.-X., {Lan-Tian} Y., {Xue-Bing} W., {Ye} L., 2000, \mnras, 317, 737

\bibitem[{{Eardley}, {Lightman} \& {Shapiro}(1975){Eardley}, {Lightman}, \&
  {Shapiro}}]{Eardley1975}
{Eardley} D.~M., {Lightman} A.~P., {Shapiro} S.~L., 1975, \apjl, 199, L153

\bibitem[{{Frank}, {King} \& {Raine}(2002){Frank}, {King}, \&
  {Raine}}]{Frank2002}
{Frank} J., {King} A., {Raine} D.~J., 2002, {Accretion Power in Astrophysics:
  Third Edition}. p. 398

\bibitem[{{Fukue}(1987)}]{Fukue87}
{Fukue} J., 1987, \pasj, 39, 309

\bibitem[{{Fukumura} \& {Tsuruta}(2004)}]{Fukumura2004}
{Fukumura} K., {Tsuruta} S., 2004, \apj, 611, 964

\bibitem[\protect\citeauthoryear{Holzer}{1977}]{Holzer_1977}
Holzer T.~E., 1977, JGR, 82, 23

\bibitem[{{Iyer}, {Nandi} \& {Mandal}(2015){Iyer}, {Nandi}, \&
  {Mandal}}]{Iyer-etal15}
{Iyer} N., {Nandi} A., {Mandal} S., 2015, \apj, 807, 108

\bibitem[\protect\citeauthoryear{Jacques}{1978}]{Jacques_1978}
Jacques S.~A., 1978, ApJ, 226, 632

\bibitem[{{Kato}, {Honma} \& {Matsumoto}(1988){Kato}, {Honma}, \&
  {Matsumoto}}]{kato1988viscous}
{Kato} S., {Honma} F., {Matsumoto} R., 1988, \mnras, 231, 37

\bibitem[{{Kato} {et~al}\mbox{.}(1993){Koto}, {Wu}, {Yang}, \&
  {Yang}}]{Kato-etal1993}
{Kato} S., {Wu} X.~B., {Yang} L.~T., {Yang} Z.~L., 1993, \mnras, 260, 317

\bibitem[\protect\citeauthoryear{Kopp \& Holzer}{1976}]{Kopp_Holzer_1976} 
Kopp R.~A., Holzer T.~E., 1976, SoPh, 49, 43

\bibitem[{{Kumar} \& {Chattopadhyay}(2014)}]{Kumar2014}
{Kumar} R., {Chattopadhyay} I., 2014, \mnras, 443, 3444

\bibitem[{{Kumar} {et~al}\mbox{.}(2013){Kumar}, {Singh}, {Chattopadhyay}, \&
  {Chakrabarti}}]{Kumar2013}
{Kumar} R., {Singh} C.~B., {Chattopadhyay} I., {Chakrabarti} S.~K., 2013,
  \mnras, 436, 2864

\bibitem[{{Kusunose} \& {Takahara}(1988)}]{kusunose1988two}
{Kusunose} M., {Takahara} F., 1988, \pasj, 40, 435

\bibitem[{{Kusunose} \& {Takahara}(1989)}]{kusunose1989two}
{Kusunose} M., {Takahara} F., 1989, \pasj, 41, 263

\bibitem[{{Landau} \& {Lifshitz}(1959)}]{Landau-Lifshitz59}
{Landau} L.~D., {Lifshitz} E.~M., 1959, {Fluid mechanics}

\bibitem[{{Lightman} \& {Eardley}(1974)}]{Lightman1974}
{Lightman} A.~P., {Eardley} D.~M., 1974, \apjl, 187, L1

\bibitem[{{Lu}, {Gu} \& {Yuan}(1999){Lu}, {Gu}, \& {Yuan}}]{Lu1999}
{Lu} J.-F., {Gu} W.-M., {Yuan} F., 1999, \apj, 523, 340

\bibitem[{{Luo} \& {Liang}(1994)}]{luo1994stability}
{Luo} C., {Liang} E.~P., 1994, \mnras, 266, 386

\bibitem[{{Mandal} \& {Chakrabarti}(2005)}]{Mandal2005}
{Mandal} S., {Chakrabarti} S.~K., 2005, \aap, 434, 839

\bibitem[{{Manmoto}, {Mineshige} \& {Kusunose}(1997){Manmoto}, {Mineshige}, \&
  {Kusunose}}]{manmoto1997spectrum}
{Manmoto} T., {Mineshige} S., {Kusunose} M., 1997, \apj, 489, 791

\bibitem[{{Matsumoto} {et~al}\mbox{.}(1984){Matsumoto}, {Kato}, {Fukue}, \&
  {Okazaki}}]{Matsumoto1984}
{Matsumoto} R., {Kato} S., {Fukue} J., {Okazaki} A.~T., 1984, \pasj, 36, 71

\bibitem[{{Misra} \& {Melia}(1996)}]{misra1996crucial}
{Misra} R., {Melia} F., 1996, \apj, 465, 869

\bibitem[{{Molteni}, {Ryu} \& {Chakrabarti}(1996){Molteni}, {Ryu}, \&
  {Chakrabarti}}]{Molteni1996a}
{Molteni} D., {Ryu} D., {Chakrabarti} S.~K., 1996, \apj, 470, 460

\bibitem[{{Molteni}, {Sponholz} \& {Chakrabarti}(1996){Molteni}, {Sponholz}, \&
  {Chakrabarti}}]{Molteni1995}
{Molteni} D., {Sponholz} H., {Chakrabarti} S.~K., 1996, \apj, 457, 805

\bibitem[{{Nakamura} {et~al}\mbox{.}(1996){Nakamura}, {Matsumoto}, {Kusunose},
  \& {Kato}}]{nakamura1996global}
{Nakamura} K.~E., {Matsumoto} R., {Kusunose} M., {Kato} S., 1996, \pasj, 48,
  761

\bibitem[{{Nandi} {et~al}\mbox{.}(2001{\natexlab{a}}){Nandi}, {Chakrabarti},
  {Vadawale}, \& {Rao}}]{Nandi_etal01a}
{Nandi} A., {Chakrabarti} S.~K., {Vadawale} S.~V., {Rao} A.~R.,
  2001{\natexlab{a}}, \aap, 380, 245

\bibitem[{{Nandi} {et~al}\mbox{.}(2012){Nandi}, {Debnath}, {Mandal}, \&
  {Chakrabarti}}]{Nandi_etal12}
{Nandi} A., {Debnath} D., {Mandal} S., {Chakrabarti} S.~K., 2012, \aap, 542,
  A56

\bibitem[{{Nandi} {et~al}\mbox{.}(2001{\natexlab{b}}){Nandi}, {Manickam},
  {Rao}, \& {Chakrabarti}}]{Nandi_etal01b}
{Nandi} A., {Manickam} S.~G., {Rao} A.~R., {Chakrabarti} S.~K.,
  2001{\natexlab{b}}, \mnras, 324, 267

\bibitem[{{Narayan} \& {Popham}(1993)}]{Narayan_pophem1993}
{Narayan} R., {Popham} R., 1993, \nat, 362, 820

\bibitem[{{Narayan} \& {Yi}(1994)}]{Narayan1994}
{Narayan} R., {Yi} I., 1994, \apjl, 428, L13

\bibitem[{{Narayan} \& {Yi}(1995)}]{Narayan-Yi95}
{Narayan} R., {Yi} I., 1995, \apj, 452, 710

\bibitem[{{Okuda} \& {Das}(2015)}]{Okuda2015}
{Okuda} T., {Das} S., 2015, \mnras, 453, 147

\bibitem[{{Paczy{\'n}sky} \& {Wiita}(1980)}]{Paczynsky_wiita1980}
{Paczy{\'n}sky} B., {Wiita} P.~J., 1980, \aap, 88, 23

\bibitem[{{Piran}(1978)}]{piran1978role}
{Piran} T., 1978, \apj, 221, 652

\bibitem[{{Pringle}(1976)}]{pringle1976thermal}
{Pringle} J.~E., 1976, \mnras, 177, 65

\bibitem[{{Radhika} \& {Nandi}(2014)}]{Radhika-Nandi14}
{Radhika} D., {Nandi} A., 2014, Advances in Space Research, 54, 1678

\bibitem[{{Radhika} {et~al}\mbox{.}(2016{\natexlab{a}}){Radhika}, {Nandi},
  {Agrawal}, \& {Mandal}}]{Radhika-etal16a}
{Radhika} D., {Nandi} A., {Agrawal} V.~K., {Mandal} S., 2016{\natexlab{a}},
  \mnras, 462, 1834

\bibitem[{{Radhika} {et~al}\mbox{.}(2016{\natexlab{b}}){Radhika}, {Nandi},
  {Agrawal}, \& {Seetha}}]{Radhika-etal16b}
{Radhika} D., {Nandi} A., {Agrawal} V.~K., {Seetha} S., 2016{\natexlab{b}},
  \mnras, 460, 4403

\bibitem[{{Rajesh} \& {Mukhopadhyay}(2010)}]{Rajesh2010}
{Rajesh} S.~R., {Mukhopadhyay} B., 2010, \mnras, 402, 961

\bibitem[{{Rybicki} \& {Lightman}(1979)}]{Rybicki1979}
{Rybicki} G.~B., {Lightman} A.~P., 1979, {Radiative processes in astrophysics}

\bibitem[{{Ryu}, {Chattopadhyay} \& {Choi}(2006){Ryu}, {Chattopadhyay}, \&
  {Choi}}]{Ryu2006}
{Ryu} D., {Chattopadhyay} I., {Choi} E., 2006, \apjs, 166, 410

\bibitem[{{Sarkar} \& {Das}(2016)}]{Sarkar-Das16}
{Sarkar} B., {Das} S., 2016, \mnras, 461, 190

\bibitem[\protect\citeauthoryear{Sarkar et al.}{2017}]{Sarkar-etal17} 
Sarkar B., Das S., Mandal S., 2017, MNRAS, in press 

\bibitem[{{Shakura} \& {Sunyaev}(1973)}]{Shakura_sunyav1973}
{Shakura} N.~I., {Sunyaev} R.~A., 1973, \aap, 24, 337

\bibitem[{{Shakura} \& {Sunyaev}(1976)}]{shakura1976}
{Shakura} N.~I., {Sunyaev} R.~A., 1976, \mnras, 175, 613

\bibitem[{{Shapiro}, {Lightman} \& {Eardley}(1976){Shapiro}, {Lightman}, \&
  {Eardley}}]{SLE1976}
{Shapiro} S.~L., {Lightman} A.~P., {Eardley} D.~M., 1976, \apj, 204, 187

\bibitem[{{Spitzer}(1965)}]{Spitzer2013}
{Spitzer} L., 1965, {Physics of fully ionized gases}

\bibitem[{{Wandel} \& {Liang}(1991)}]{wandel1991hybrid}
{Wandel} A., {Liang} E.~P., 1991, \apj, 380, 84

\bibitem[{{White} \& {Lightman}(1989)}]{white1989hot}
{White} T.~R., {Lightman} A.~P., 1989, \apj, 340, 1024

\bibitem[{{Wu}(1997)}]{wu1997stability}
{Wu} X.-B., 1997, \mnras, 292, 113

\end{thebibliography}



\appendix


\label{lastpage}
\end{document}